\documentclass[11pt,draftcls,peerreview,onecolumn]{IEEEtran}

\usepackage[dvips]{graphics}
\usepackage{graphicx}
\usepackage{times}
\usepackage{amsmath}
\usepackage{amssymb}
\usepackage{fancybox}
\newtheorem{property}{\bf Property}
\newtheorem{lemma}{\bf Lemma}
\newtheorem{theorem}{\bf Theorem}

\newtheorem{algorithm}{\bf Algorithm}

\newcommand{\bee}{\begin{eqnarray}}
\newcommand{\eee}{\end{eqnarray}}
\newcommand{\be}{\begin{equation}}
\newcommand{\ee}{\end{equation}}

\newcommand{\al}[1]{\begin{align} #1 \end{align}}

\newcommand{\equ}[1]{\begin{equation} #1 \end{equation}}

\newcommand{\mb}{\mathbf}

\newcommand{\nnb}{\nonumber}

\newcommand{\qa}{{\bf a}}

\newcommand{\qh}{{\bf h}}

\newcommand{\qu}{{\bf u}}
\newcommand{\qv}{{\bf v}}
\newcommand{\qw}{{\bf w}}
\newcommand{\qx}{{\bf x}}

\newcommand{\qz}{{\bf z}}

\newcommand{\qA}{{\bf A}}
\newcommand{\qB}{{\bf B}}
\newcommand{\qC}{{\bf C}}
\newcommand{\qD}{{\bf D}}

\newcommand{\qF}{{\bf F}}
\newcommand{\qG}{{\bf G}}

\newcommand{\qI}{{\bf I}}
\newcommand{\qJ}{{\bf J}}
\newcommand{\qK}{{\bf K}}

\newcommand{\qQ}{{\bf Q}}
\newcommand{\qR}{{\bf R}}
\newcommand{\qS}{{\bf S}}

\newcommand{\qX}{{\bf X}}

\begin{document}
%
\title{On Cooperative Beamforming Based on Second-Order Statistics of Channel State Information}

\author{\IEEEauthorblockN{Jiangyuan Li, Athina P. Petropulu, H. Vincent Poor$^*$  }\\
\IEEEauthorblockA{Department of Electrical and Computer Engineering,
Drexel University, Philadelphia, PA 19104\\
$^*$School of Engineering and Applied Science,
          Princeton University, Princeton, NJ 08544
}
}

\maketitle

\begin{abstract}
\footnote{This research was supported in part by the Office of Naval Research under Grants ONR-N-00010710500, N-00014-09-1-0342 and in part by the
National Science Foundation under Grants CNS-0905425, CNS-09-05398.}
Cooperative beamforming in relay networks is considered,
in which a source transmits to its  destination with the help of a set of cooperating  nodes.
The source first transmits locally. The cooperating nodes that receive the source signal retransmit a weighted version of it in an
amplify-and-forward (AF) fashion.
Assuming knowledge of the second-order statistics of the channel state information, beamforming weights are determined so that the signal-to-noise ratio (SNR) at the destination is maximized subject to two different power constraints, i.e., a total (source and relay) power constraint,   and
individual relay power constraints.
For the former constraint, the original problem is transformed into
a problem of one variable, which can be solved via Newton's method.
For the latter constraint,  the original problem is transformed into a homogeneous quadratically constrained quadratic programming (QCQP) problem. In this case, it is shown that
 when the number of relays does not exceed three the global solution can always be constructed via semidefinite programming (SDP)
relaxation and the matrix rank-one decomposition technique. For the cases in which the SDP relaxation does not generate a rank one solution,
 two methods are proposed to solve the problem: the first one is
based on the coordinate descent method, and the second one
 transforms the QCQP problem into an infinity norm maximization problem in which a smooth finite norm approximation  can lead to the solution using the
augmented Lagrangian method.
\end{abstract}



\section{Introduction}

Cooperative beamforming (CB), also called  distributed
beamforming has attracted considerable research interest recently, due to its
potential for improving communication reliability.
One form of distributed beamforming, the so-called distributed transmit beamforming, is a form
of cooperative communications in which a network of multiple transmitters cooperate to transmit a common
message coherently to a Base Station (BS).
The distributed transmit beamforming can provide energy efficiency and reasonable directional gain for ad hoc sensor networks \cite{Poor1}, \cite{Poor3}.
The challenges and recent progress of distributed transmit beamforming are discussed in \cite{Poor2}.
Another form of distributed beamforming is the distributed relay beamforming, in which
a set of cooperating nodes act as a virtual antenna array and
adjust their transmission weights to form a beam to the destination. This can
 result in diversity gains similar to those of  multiple-antenna systems \cite{Jing1}, \cite{Luo}.
Various effective cooperation schemes have been proposed in the literature, such as amplify-and-forward (AF),
decode-and-forward (DF) \cite{Tse}, coded-cooperation \cite{Janani}, and
compress-and-forward \cite{Kramer}.
The AF protocol, due to its simplicity, is of particular interest \cite{Luo}.

In distributed relay beamforming, the objective is to determine  source power and  beamforming weights according to some optimality criterion.
Existing results for this problem can be classified into those that rely on channel state information (CSI) availability at the relays \cite{Jing1}, \cite{Dong}, \cite{Veria},
and those that allow for channel uncertainly, i.e.,  that rely on statistics of CSI, such as the covariance of channel coefficients, or imperfect CSI feedback \cite{Luo}, \cite{Siavash}, \cite{Jing2}, as opposed to explicit CSI.
The latter class of techniques is particularly important because CSI is never perfectly known at the transmitter.
This work picks up on some important results presented in   \cite{Luo},
 in which a source transmits a signal to a destination with the assistance of a set of AF relay nodes
In \cite{Luo}, the problem of obtaining the
beamforming weights  so that the signal-to-noise ratio (SNR) at the destination is
maximized subject to certain power constraints is considered, i.e., individual relay power constraints and a total power relay constraint.
For  the case of  individual relay power constraints, a semidefinite programming (SDP) relaxation
plus bisection search technique was proposed in \cite{Luo}. When the SDP relaxation generates a rank-one solution, then this is the exact solution of the original problem;
otherwise, the exact solution cannot be guaranteed, and the authors of  \cite{Luo} proposed a Gaussian random procedure (GRP) to search for an approximate solution based on the SDP relaxation solution. However, GRP is time-consuming and sometimes ineffective.

In this paper, we investigate the same scenario as in \cite{Luo}, i.e.,  cooperative beamforming under the assumption that the second-order statistics
of the channel state information (CSI) are available.
The beamforming weights are determined so that the SNR at the destination is maximized subject to two  different power constraints: (i)
  a total (source plus relay) power constraint, and (ii) individual relay constraints.
The differences of this work as compared to \cite{Luo}, are the following.
\begin{itemize}
  \item  Our first kind of power constraint  includes the source power as well as the power of the relays. In a wireless network all nodes have power constraints, therefore, placing a constraint on the source is more realistic. However, this results in a more difficult optimization problem. A similar constraint was also used in  \cite{Dong2}.
 For this case, we transform the original problem  into
a problem of one variable, which can then be solved via Newton's method.
  \item The second kind of power constraint is exactly the same as that of \cite{Luo}, but our work contributes new results and more efficient algorithms to  reach the solution. In particular,
  \begin{itemize}
  \item
  We show that when the number of relays does not exceed three, the global solution can always be constructed via SDP relaxation and the matrix rank-one decomposition technique.
  \item For the case in which the SDP relaxation solution has rank greater than one, we propose two methods to obtain an approximate solution that is more
   effective than the Gaussian random procedure employed  in \cite{Luo}.
    The first method is based on the coordinate descent method.
The second method transforms the original problem into an infinity norm maximization problem, for which  a smooth finite norm approximation results in a solution using the
 augmented Lagrangian method.
  \end{itemize}
  \item For both types of constraints, we obtain exact solutions for  the special cases in which the channel coefficients between different node pairs are uncorrelated and follow  a Rayleigh fading model. These cases were not discussed in \cite{Luo}.

\end{itemize}

The remainder of the paper is organized as follows. The mathematical model is introduced in \S\ref{Sec:SysModel}.
In \S\ref{Sec:SNRmaxTotalPower}, the SNR maximization subject to a total power constraint is presented.
The SNR maximization subject to individual relay power constraints is developed in Section \S\ref{Sec:SNRmaxIndiRelayPower}.
Numerical results are presented in \S\ref{Sec:Sim} to illustrate the proposed algorithms.
Finally, \S\ref{Sec:Conclu} provides concluding remarks.

\subsection{Notation}

Upper case and lower case bold symbols denote matrices and vectors, respectively.
Superscripts $\ast$, $T$ and $\dagger$
denote respectively conjugate, transposition and conjugate transposition.
$|\cdot|$ denotes the amplitude of a complex number.
$\mathrm{det}(\qA)$ and
$\mathrm{Tr}({\mb A})$ denote  determinant and trace of matrix $\mb A$, respectively.
$\lambda_{\min}(\qA)$ and $\lambda_{\max}(\qA)$ denote the smallest and largest eigenvalues of $\qA$, respectively.
${\mb A}\succeq 0$ and ${\mb A}\succ 0$ mean that matrix ${\mb A}$ is
Hermitian positive semidefinite, and positive definite, respectively.
$\qA\succeq \qB$ denotes that $\qA-\qB$ is a positive semidefinite matrix.
$\mathrm{rank}(\qA)$ denotes the rank of matrix $\qA$.
$\mathrm{diag}(\qv)$ denotes a diagonal matrix with diagonal entries consisting of the elements of $\qv$.
$\|\qa\|$ denotes Euclidean norm of vector $\qa$.
$\qI_n$ denotes the identity matrix of order $n$
(the subscript is dropped when the dimension is obvious).
$\mathbb{E}(\cdot)$ denotes expectation.

\begin{figure}[hbtp]
\centering
\includegraphics[width=3.5in]{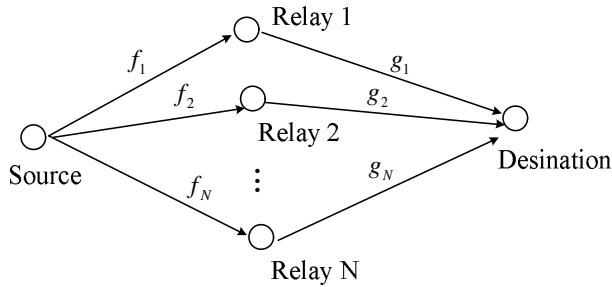}
\caption{System model.}
\label{fig:1}
\end{figure}

\section{System Model and Problem Statement}\label{Sec:SysModel}

The system model is the same as in \cite{Luo} and is  depicted in Fig. \ref{fig:1}. It consists of a source node, a destination node and $N$ relay nodes,
each node equipped with a single antenna.
The source transmits signals to the destination with the help of relay nodes.
We assume that the direct link between the source and destination is very weak and thus ignored.
The channel gains from the source to the $i$th relay, and from the $i$th relay to the destination, are denoted respectively by $f_i$ and $g_i$.

Communication between source and destination occurs in two stages (slots). During the first stage, the source broadcasts its signal to
the relays. During the second stage, the relays working in AF fashion transmit a weighted version of the signal
that they received during the first stage.
Let $\sqrt{P_s}\,s$ be the source signal,  where $P_s$ is the source transmit power and $s$ is the information symbol with $\mathbb{E}(|s|^2)=1$. The received signal at the $i$th relay is given by
\equ{
x_i=\sqrt{P_s}\, f_i s+v_i
}
where $v_i$ represents the noise at the $i$th relay having zero mean and variance $\sigma^2$.
The $i$th relay weights the received signal and transmits $z_i=w_i x_i$
where $w_i$ is the weight. The received signal at the destination equals
\equ{
y=\sum_{i=1}^N g_i z_i + \nu=\sqrt{P_s}\sum_{i=1}^N w_if_i g_i s+\sum_{i=1}^N w_ig_iv_i+\nu\label{Rece}
}
where $\nu$ is the noise at the destination having zero mean and variance $\sigma^2$.

Let us assume that the second-order statistics of the channel gains $f_i$'s
and $g_i$'s are known. We also assume that $f_i$ and $g_j$, $\forall i, j$ are statistically independent. Define
\al{
\qw &= [w_1, \cdots, w_N]^T, \nnb\\
\qh &= [f_1g_1,\cdots,f_Ng_N]^T, \nnb\\
\mb{g} &= [g_1,\cdots, g_N]^T, \nnb\\
\qR &= \mathbb{E}\{\qh\qh^\dagger\}, \nnb\\
\qQ &= \mathbb{E}\{\mb{g}\mb{g}^\dagger\},\nnb\\
\mathrm{and}\ \qD &= \mathrm{diag}(\mathbb{E}\{|f_1|^2\}, \cdots, \mathbb{E}\{|f_N|^2\}).
}
In general, $\qQ$ and $\qR$ are full matrices.
In case of uncorrelated Rayleigh fading, in holds that $\mathbb{E}(f_i^\ast f_j)=0$, and $\mathbb{E}(g_i^\ast g_j)=0$, $\forall i\ne j$,
in which case $\qR$ and $\qQ$ both are diagonal.

From (\ref{Rece}), the signal component power is given by
\equ{
P_d=\mathbb{E}\left\{\bigg\vert\sqrt{P_s}\sum_{i=1}^N w_if_i g_i s\bigg\vert^2\right\}=P_s\qw^\dagger\qR\qw\label{Pd}
}
and the total noise power $P_n$ equals
\equ{
P_n=\mathbb{E}\left\{\bigg\vert\sum_{i=1}^N w_ig_iv_i+\nu\bigg\vert^2\right\}=\sigma^2+\sigma^2\qw^\dagger\qQ\qw.
}
The SNR at the destination is given by
\equ{
\Gamma_d=\frac{P_d}{P_n}=\frac{P_s}{\sigma^2}\frac{\qw^\dagger\qR\qw}{1+\qw^\dagger\qQ\qw}.\label{SNR}
}

The total relay transmit  power and transmit power at the $i$th relay are respectively given by
\al{
P_r&=\sum_{i=1}^N \mathbb{E}\{|z_i|^2\}=P_s\qw^\dagger\qD\qw + \sigma^2\qw^\dagger\qw\\
P_{r,i}&=\mathbb{E}\{|z_i|^2\}=(P_s D_{ii}+\sigma^2) |w_i|^2
}
where $D_{ii}$ is the $(i,i)$th entry of $\qD$.

Our goal in this paper is to determine the beamforming weights $w_i$'s such that
$\Gamma_d$ is maximized subject to certain power constraints.
In this paper, we consider two kinds of power constraints.
The first kind corresponds to the case in which the total power of the source and all relays is constrained, i.e.,
\equ{
P_s+P_r\le P_0.\label{TotPowCons}
}
where $P_0$ is the maximum allowable total transmit power of the source and all relays.
The second kind is the individual relay power constraints in which
each relay node is restricted in its transmit power, i.e,
\equ{
P_{r,i}\le P_i\label{IndRelPowCons}
}
where $P_i$ is the maximum allowable transmit power of the $i$th
relay.

\section{SNR Maximization Under Total Power Constraint}\label{Sec:SNRmaxTotalPower}

From (\ref{SNR}) and (\ref{TotPowCons}), the SNR maximization problem subject to a total power constraint is expressed as
\al{
&\max_{P_s, \qw}\ \frac{P_s}{\sigma^2}\frac{\qw^\dagger\qR\qw}{1+\qw^\dagger\qQ\qw}\label{totSNRmax}\\
&\mathrm{s.t.} \quad P_s+P_s\qw^\dagger\qD\qw + \sigma^2\qw^\dagger\qw \le P_0.\nnb
}
We give the following lemma, the proof of which can be found in   Appendix \ref{ProofLem:totSNRmax}.\\

\begin{lemma}\label{Lem:totSNRmax}
{\em Let $P_s^\circ$ be the solution of the following
\al{
&\max_{P_s}\ \frac{P_s}{\sigma^2}\frac{P_0-P_s}{
\lambda_{\min}(P_s\qS_1+(P_0-P_s)\qS_2)}\label{totSNRmax4}\\
&\mathrm{s.t.} \quad 0\le P_s\le P_0\nnb
}
where
\al{
\qS_1&=\qR^{-1/2}\qD\qR^{-1/2}+(\sigma^2/P_0)\qR^{-1},\label{S1}\\
\mathrm{and} \ \qS_2&=\qR^{-1/2}\qQ\qR^{-1/2}+(\sigma^2/P_0)\qR^{-1}.\label{S2}
}
Let $\qw^\circ$ be the eigenvector associated with the smallest eigenvalue of $P_s^\circ\qS_1+(P_0-P_s^\circ)\qS_2$.
Then $(P_s^\circ, \qw^\circ)$ is the solution to the problem of (\ref{totSNRmax}).
}
\end{lemma}
{\em Remarks}: Here we assume that $\qR\succ 0$. If $\qR\nsucc 0$, the methodology is similar.
In fact, from Appendix \ref{ProofLem:totSNRmax}, the problem of (\ref{totSNRmax}) is also equivalent to
\al{
&\max_{P_s}\ P_s(P_0-P_s)\lambda_{\max}\left([P_s\qD+\sigma^2\qI+(P_0-P_s)\qQ]^{-\frac{1}{2}}\qR[P_s\qD+\sigma^2\qI+(P_0-P_s)\qQ]^{-\frac{1}{2}}\right)\\
&\mathrm{s.t.} \quad 0\le P_s\le P_0.\nnb
}
A similar procedure can be used to solve the above problem.

Let us normalize $P_s$ by letting $x=P_s/P_0$, $0\le x\le 1$.
With this, the problem of (\ref{totSNRmax4}) is equivalent to
\al{
&\max_{x}\ \frac{P_0}{\sigma^2}\frac{x(1-x)}{
\lambda_{\min}(x\qS_1+(1-x)\qS_2)}\label{totSNRmax4a}\\
&\mathrm{s.t.} \quad 0\le x\le 1.\nnb
}

\subsection{$\qS_1$ and $\qS_2$  are both diagonal}\label{diagonalTotal}

In case of uncorrelated Rayleigh fading,  $\qR$ and $\qQ$ are diagonal matrices. Then, $\qS_1$ and $\qS_2$ are both diagonal, and as it will be shown next the exact solution can be obtained analytically.

By denoting the $(k,k)$-th entry of $\qS_1$ and $\qS_2$ as $a_k$ and $b_k$, respectively, the problem of (\ref{totSNRmax4a}) becomes
\al{
&\min_{0<x<1}\ \lambda_{\min}\left(\frac{1}{1-x}\qS_1+\frac{1}{x}\qS_2\right) \nnb \\
=\ &\min_{0<x<1}\ \min_{k=1,\cdots,N}\ \bigg\{ \frac{a_k}{1-x}+\frac{b_k}{x}\bigg\}\nnb\\
=\ &\min_{k=1,\cdots,N}\ \min_{0<x<1}\ \bigg\{ \frac{a_k}{1-x}+\frac{b_k}{x}\bigg\}\nnb\\
=\ &\min_{k=1,\cdots,N}\ (\sqrt{a_k}+\sqrt{b_k})^2\nnb\\
=\ & (\sqrt{a_{k_0}}+\sqrt{b_{k_0}})^2.
}
The above minimum is attained for
\equ{
x=\frac{\sqrt{b_{k_0}}}{\sqrt{a_{k_0}}+\sqrt{b_{k_0}}}
}
where
\equ{
k_0=\arg \min_{k=1,\cdots,N}\ (\sqrt{a_k}+\sqrt{b_k})^2.
}


\subsection{$\qS_1$ or $\qS_2$ is not diagonal}

\medskip
\begin{lemma}\label{Lem:Opt_x_interval}
{\em The optimal $x$ of (\ref{totSNRmax4a}) lies in $[x_l, x_u]$ where
\al{
x_l&=\frac{\sqrt{c}}{1+\sqrt{c}},\\
\mathrm{and}\ x_u&=\frac{\sqrt{d}}{1+\sqrt{d}}
}
where $c=\lambda_{\min}(\qS_1^{-1/2}\qS_2\qS_1^{-1/2})$, and $d=\lambda_{\max}(\qS_1^{-1/2}\qS_2\qS_1^{-1/2})$.
}
\end{lemma}
The proof is given in Appendix \ref{ProofLem:Opt_x_interval}.
\medskip

From Lemma \ref{Lem:Opt_x_interval}, to solve the problem of (\ref{totSNRmax4a}) is equivalent to solving the problem of
\al{
&\min_{x}\ \lambda_{\min}\left(\frac{1}{1-x}\qS_1+\frac{1}{x}\qS_2\right) \label{totSNRmax4b}\\
&\mathrm{s.t.} \quad x_l\le x \le x_u.\nnb
}

The objective in (\ref{totSNRmax4b}) is in general not a convex function over $[x_l, x_u]$.
We will use Newton's method to search for the stationary points.
Let us start by denoting
\equ{
\qG(x)=\frac{1}{1-x}\qS_1+\frac{1}{x}\qS_2, \ x\in [x_l, x_u].
}
Note that $\qG(x)$ depends smoothly on $x\in [x_l, x_u]$ as any order derivative of $\qG(x)$ exists.
We assume that $\qG(x)$ has a simple spectrum for $x\in [x_l, x_u]$.
This is a reasonable assumption for general $\qS_1$ and $\qS_2$ (see \cite{Still}, \cite[\S 4]{Tao}).
Under this assumption, $\lambda_{\min}(\qG(x))$ also depends smoothly on $x\in [x_l, x_u]$ \cite{Still}.
First- and second- order necessary conditions for $x$ to be a local minimizer are respectively \cite[Theorem 2.2, 2.3]{Nocedal}
\al{
&\frac{\mathrm{d}}{\mathrm{d} x}\lambda_{\min}(\qG(x))=0,\label{firstNessCond}\\
\mathrm{and}\ &\frac{\mathrm{d}^2 }{\mathrm{d} x^2}\lambda_{\min}(\qG(x))\ge 0.\label{secondNessCond}
}
If (\ref{secondNessCond}) holds with strict inequality, then $x$ is a strict local minimizer \cite[Theorem 2.4]{Nocedal}.
In Newton's method, the $(k+1)$th iteration is given by \cite[Ch. 3]{Nocedal}
\equ{
x_{k+1}=x_k-\alpha_k\frac{\frac{\mathrm{d}}{\mathrm{d} x}\lambda_{\min}(\qG(x))}
{\frac{\mathrm{d}^2 }{\mathrm{d} x^2}\lambda_{\min}(\qG(x))}, \ k=0,1,\cdots\label{iter}
}
where $\alpha_k>0$ is chosen such that $x_{k+1}$ does not exceed $[x_l, x_u]$,
and otherwise, $\alpha_k \leftarrow \alpha_k/2$.

In the iteration expression (\ref{iter}), we need to calculate the first- and second- order derivatives of $\lambda_{\min}(\qG(x))$.
Let $\qu_0(x)$ be the eigenvector associated with $\lambda_{\min}(\qG(x))$.
Let $\qu_k(x)$, $k=1,\cdots,N-1$ be the eigenvectors associated with the other eigenvalues $\lambda_k(x)$ of $\qG(x)$, respectively,
where $\lambda_1(x)>\cdots>\lambda_{N-1}(x)>\lambda_{\min}(\qG(x))$.
The first- and second- order derivatives of $\lambda_{\min}(\qG(x))$ (the so-called {\em Hadamard first
variation formula} and {\em Hadamard second variation formula} \cite[\S 4]{Tao}) are respectively given by \cite{VanDerAa}, \cite{Overton}
\al{
\frac{\mathrm{d}}{\mathrm{d} x}\lambda_{\min}(\qG(x))&=\qu_0(x)^\dagger\frac{\mathrm{d} \qG(x)}{\mathrm{d} x}\qu_0(x),\\
\mathrm{and}\ \frac{\mathrm{d}^2 }{\mathrm{d} x^2}\lambda_{\min}(\qG(x))&=\qu_0(x)^\dagger\frac{\mathrm{d}^2 \qG(x)}{\mathrm{d} x^2}\qu_0(x)
-\sum_{j=1}^{N-1}\frac{2\big\vert\qu_j(x)^\dagger\frac{\mathrm{d} \qG(x)}{\mathrm{d} x}\qu_0(x)\big\vert^2}{\lambda_j(x)-\lambda_{\min}(\qG(x))}
}
where
\al{
\frac{\mathrm{d} \qG(x)}{\mathrm{d} x}&=\frac{1}{(1-x)^2}\qS_1-\frac{1}{x^2}\qS_2,\\
\mathrm{and}\ \frac{\mathrm{d}^2 \qG(x)}{\mathrm{d} x^2}&=\frac{2}{(1-x)^3}\qS_1+\frac{2}{x^3}\qS_2.
}

\section{SNR Maximization Under Individual Relay Power Constraints}\label{Sec:SNRmaxIndiRelayPower}

From (\ref{SNR}) and (\ref{IndRelPowCons}), the SNR maximization problem subject to individual relay power constraints is expressed as
\al{
&\max_{\qw}\ \frac{P_s}{\sigma^2}\frac{\qw^\dagger\qR\qw}{1+\qw^\dagger\qQ\qw}\label{indSNRmax1}\\
&\mathrm{s.t.} \ \ (P_s D_{kk}+\sigma^2) |w_k|^2\le P_k, \ k\in I\nnb
}
where $I=\{1,2,\cdots,N\}$. The problem of (\ref{indSNRmax1}) belongs to the class of quadratically constrained fractional programs. In \cite{Luo}, this problem was analyzed and an SDP relaxation plus bisection search technique
was proposed. Here, we first consider the case of uncorrelated Rayleigh fading, and show that an exact solution can be obtained. Then, for the general fading case, we propose two methods that are more efficient than the search method of \cite{Luo}.
As it will be shown in the simulations section, the random search approach, in addition to being time consuming,  can result in a noticeable performance gap as compared to the proposed approaches.

\subsection{$\qR$ and $\qQ$  are both diagonal}\label{Subsec:indiDiag}

By using the Dinkelbach-type method \cite{Dinkelbach}, we introduce the following function:
\al{
F(t)=&\max_{\qw}\ \bigg[f(t, \qw)=\frac{P_s}{\sigma^2}\qw^\dagger\qR\qw - t(1+\qw^\dagger\qQ\qw) \bigg]\label{indSNRmax2}\\
&\mathrm{s.t.}\ \ (P_s D_{kk}+\sigma^2) |w_k|^2\le P_k, \ k\in I.\nnb
}
The relation between $F(t)$ and the problem of (\ref{indSNRmax1}) is given in the following property \cite{Dinkelbach}.
\medskip
\begin{property}\label{prop:F}
\mbox{}
{\em
\begin{itemize}
  \item [(i)] $F(t)$ is strictly decreasing, and $F(t)=0$ has a unique root, say $t^\star$;
  \item [(ii)] Let $\qw^\star$ be the solution of (\ref{indSNRmax2}) corresponding to $t^\star$.
Then $\qw^\star$ is also the solution of (\ref{indSNRmax1}) with the largest objective value $t^\star$ exactly.
\end{itemize}
}
\end{property}
\medskip


According to Property \ref{prop:F}, we aim to find $t^\star$ and the associated $\qw^\star$,
which is also the solution of (\ref{indSNRmax1}). To this end, by denoting the $(k,k)$th entry of $\qR$, $\qQ$
as $r_k$, $q_k$, respectively, we rewrite
\equ{
f(t, \qw)=-t+\sum_{k=1}^N\bigg(\frac{P_s}{\sigma^2}r_k-tq_k\bigg)|w_k|^2
}
to get that
\equ{
F(t) = -t+\sum_{k=1}^N\frac{P_k}{P_s D_{kk}+\sigma^2}\,\varphi\bigg(\frac{P_s}{\sigma^2}r_k-tq_k\bigg)\label{F_t}
}
associated with the optimal
\equ{
|w_k|^2=\left\{\begin{array}{cl}
                 \frac{P_k}{P_s D_{kk}+\sigma^2} & \frac{P_s}{\sigma^2}r_k-tq_k > 0 \\
                 0 & \mathrm{otherwise}
               \end{array}
\right.\label{Opt_w}
}
where
\equ{
\varphi(x)\triangleq\left\{\begin{array}{cl}
                    x & x>0 \\
                    0 & \mathrm{otherwise}.
                  \end{array}
\right.
}

To find the root of $F(t)=0$, let us denote
\equ{
t_k=\frac{P_sr_k}{\sigma^2q_k}, \ k=1,\cdots,N
}
and their rearrangement $\tilde{t}_1 < \tilde{t}_2 < \cdots < \tilde{t}_N$ corresponding to  $\tilde{r}_k$, $\tilde{q}_k$, $\tilde{P}_k$, and $\tilde{D}_{kk}$, respectively.
With these, we rewrite (\ref{F_t}) as
\equ{
F(t) = -t+\sum_{k=1}^N\frac{\tilde{P}_k}{P_s \tilde{D}_{kk}+\sigma^2}\,\varphi\bigg(\frac{P_s}{\sigma^2}\tilde{r}_k-t\tilde{q}_k\bigg).\label{F_t_2}
}
Note that $F(0)>0$ and $F(\tilde{t}_N)=-\tilde{t}_N < 0$. Thus, it follows from Property \ref{prop:F} that $ 0<t^\star < \tilde{t}_N$.
The root $t^\star$ is determined based on the following theorem,  the proof of which is given in Appendix \ref{ProofTheo:t_star}.
\medskip
\begin{theorem}\label{Theo:t_star}
{\em If $F(\tilde{t}_{k_0})= 0$ for an integer $k_0$, then $t^\star=t_{k_0}$. Otherwise, let $k_0$ be the smallest integer such that $F(\tilde{t}_{k_0})< 0$. Then
\equ{
t^\star = \bigg(1+\sum_{k=k_0}^N\frac{\tilde{P}_k \tilde{q}_k}{P_s \tilde{D}_{kk}+\sigma^2}\bigg)^{-1}
\sum_{k=k_0}^N \frac{\tilde{P}_k P_s \tilde{r}_k}{(P_s \tilde{D}_{kk}+\sigma^2)\sigma^2}.
}
}
\end{theorem}
Once $t^\star$ is obtained, we can obtain $\qw^\star$ from (\ref{Opt_w}).

\subsection{$\qR$ or $\qQ$ is not diagonal}\label{Sec:IndiNondiag}

\subsubsection{Equivalent QCQP and SDP relaxation}

The problem of (\ref{indSNRmax1}) is equivalent ({\em up to scaling}) to a QCQP, as stated in the following lemma.
The proof of the lemma is given in Appendix \ref{ProofLem:QCQP}.
\medskip
\begin{lemma}\label{Lem:QCQP}
{\em Let $\qw^\circ$ be the solution of the following homogeneous QCQP problem:
\al{
&\max_{\qw}\ \qw^\dagger\qR\qw\label{indNotDiagSNRmax2f}\\
&\mathrm{s.t.}\quad \qw^\dagger\qA_k\qw\le 1, \ k\in I\nnb
}
where
\equ{
\qA_k=\frac{P_s D_{kk}+\sigma^2}{P_k}\qJ_k+\qQ
}
and $\qJ_k$ is a matrix with all zero entries except for the $(k,k)$th entry one. Let
\equ{
\eta=\max_{k\in I}\ \frac{P_s D_{kk}+\sigma^2}{P_k}{\qw^\circ}^\dagger\qJ_k\qw^\circ.
}
Then $\frac{1}{\sqrt{\eta}}\qw^\circ$ is the solution to the problem of (\ref{indSNRmax1}).
}
\end{lemma}
{\em Remarks}: In fact, Lemma \ref{Lem:QCQP} states that the QCQP of (\ref{indNotDiagSNRmax2f}) and the problem of (\ref{indSNRmax1})
are equivalent {\em up to scaling}.
\medskip

Note that the constraint in (\ref{indNotDiagSNRmax2f}) is convex but the objective is concave.
Thus, the problem of (\ref{indNotDiagSNRmax2f}) is not a convex problem. In fact, this problem belongs to
the class of problems involving maximization of convex functions over a convex set \cite{Zalinescu}.

The SDP relaxation is a popular method for QCQP problems. Let $\qX=\qw\qw^\dagger$,
and we can write $\qw^\dagger\qR\qw=\mathrm{Tr}(\qR\qX)$,
$\qw^\dagger\qA_k\qw=\mathrm{Tr}(\qA_k\qX)$.
With this, we can rewrite the problem of (\ref{indNotDiagSNRmax2f}) as
\al{
&\min_{\qX}\ -\mathrm{Tr}(\qR\qX) \label{indNotDiagSNRmax2-SDP1}\\
&\mathrm{s.t.}\quad \mathrm{Tr}(\qA_k\qX)\le 1, \ k\in I\nnb\\
&\quad\quad\ \qX \succeq 0,\nnb\\
&\quad\quad\ \mathrm{rank}(\qX)=1.\nnb
}
Dropping the non-convex constraint $\mathrm{rank}(\qX)=1$, we obtain the SDP relaxation \cite{Boyd}
\al{
&\min_{\qX}\ -\mathrm{Tr}(\qR\qX) \label{indNotDiagSNRmax2-SDP2}\\
&\mathrm{s.t.}\quad \mathrm{Tr}(\qA_k\qX)\le 1, \ k\in I\nnb\\
&\quad\quad\ \qX \succeq 0.\nnb
}
The SDP of (\ref{indNotDiagSNRmax2-SDP2}) a convex problem which can be effectively solved by CVX software \cite{Grant}.
Let $\qX^\star$ be such a solution.
Obviously, if $\qX^\star$ has rank one, then it is the solution to the problem of (\ref{indNotDiagSNRmax2-SDP1})
and hence generates the solution to the problem of (\ref{indNotDiagSNRmax2f}).
Otherwise, a search technique may be used to obtain the suboptimal solution of the original problem, e.g., the Gaussian random procedure (GRP) \cite{Luo}.
For general $\qR$ and $\qQ$, the solution $\qX^\star$ from  CVX software does not necessarily have rank one
(in fact, for general $\qR$ and $\qQ$ matrices, the SDP of (\ref{indNotDiagSNRmax2-SDP2}) does not necessarily
have a rank one solution).
Some examples on the above claim will be given in the simulation section below.

The SDP relaxation problem of (\ref{indNotDiagSNRmax2-SDP2}) has several advantages as compared to
the SDP relaxation of \cite{Luo}.
First, it obtains the same objective value while avoiding
the bisection search.
Second, for $N=2, 3$, it attains the global optimal solution
in polynomial time.
In other words, for $N=2, 3$, one can ensure that the problem of (\ref{indNotDiagSNRmax2-SDP2}) has a rank one solution.
Moreover, one can construct a rank one solution from any non rank one $\qX^\star$ in polynomial time.
In fact, for $N=2, 3$, the problem has been solved  using the complex matrix rank-one
decomposition \cite[Theorem 2.1]{Huang}, as stated in the following theorem.
\medskip
\begin{theorem}\label{Theo:tightSDP}
{\em For $N=2, 3$, the problem of (\ref{indNotDiagSNRmax2-SDP2}) has a rank one solution.
Let $\qX^\star$ be any one of the solutions. If $\qX^\star$ has a rank greater than one,
one can construct a rank one solution from $\qX^\star$ in polynomial time by using
the complex matrix rank one decomposition.
}
\end{theorem}
\medskip

For the case in which the solution $\qX^\star$ from the CVX software has a rank greater than one, the GRP
can  used, although it is in general time-consuming and sometimes ineffective.
In the following, we give two more effective methods for that case.\\

\subsubsection{Coordinate descent method}\label{Sec:AlterativeMin}

If the solution $\qX^\star$ from CVX software has  rank greater than one we can use  the
coordinate descent method \cite[\S8.9]{Luenberger}, \cite[\S2.7]{Bertsekas}, \cite{Tseng}, \cite{Leeuw} to directly deal with the original problem of (\ref{indSNRmax1}).
Note that the constraints of  the problem of (\ref{indSNRmax1})
are some bounds for the elements of $\qw$, i.e., a Cartesian product of some closed convex sets (see \cite[\S2.7]{Bertsekas}).
The idea behind the coordinate descent method is the following.
At each iteration, the objective is minimized with respect to one element of $\qw$ while keeping the other elements fixed.
The method is particularly attractive when the subproblem is easy to solve (e.g., there is a closed form solution) and also satisfies
 certain condition for convergence
\cite[Proposition 2.7.1]{Bertsekas}, \cite[Theorem 4.1]{Tseng}, \cite[\S6]{Leeuw}.
The coordinate descent algorithm applied to our problem is as follows.
\medskip
\begin{algorithm}\label{algo:CDM}
\mbox{}
{\em
\begin{enumerate}
  \item [1)] Set $\varepsilon=10^{-3}$; Choose an initial point $\qw^0$; Set $k=0$.
  \item [2)] For $p=1:N$, determine the optimal $p$th element while keeping the other elements fixed. This
   results in $\qw_p^k$;
  \item [3)] $\qw^{k+1} = \qw_N^k$;
  \item [4)] If $\frac{\|\qw^{k+1}-\qw^k\|}{\|\qw^k\|} < \varepsilon$, stop;
  \item [5)] $k=k+1$; Go to 2).
\end{enumerate}
}
\end{algorithm}
\medskip

In the following, we show that the subproblem stated in Step 2 has a closed form solution (see Theorem \ref{Theo:AlterMin}) and also study its
convergence to a stationary
point (see Theorem \ref{GlobConverg}).

 It is easy to verify that
minimizing the objective with respect to the $k$th element of $\qw$ while keeping the other elements fixed
leads to the following optimization problem:
\al{
&\max_y\ \frac{a_1 |y|^2 + b_1 y + b_1^\ast y^\ast + c_1}{a_2 |y|^2 + b_2 y + b_2^\ast y^\ast + c_2}\label{indSNRmaxAlterMin}\\
&\mathrm{s.t.} \ \  |y|\le \beta\nnb
}
where $\beta = \sqrt{P_k/(P_s D_{kk}+\sigma^2)}$, $a_1 = R_{kk}$, $a_2 = Q_{kk}$ and $b_1, b_2, c_1, c_2$ can be inferred from (\ref{indSNRmax1}).
For example, when $k=1$, let $\qw=[y, \widetilde{\qw}^T]^T$ and
\equ{
\qQ = \left(
        \begin{array}{cc}
          Q_{11} & \mb{l}_1^\dagger \\
          \mb{l}_1 & \qQ_1 \\
        \end{array}
      \right), \
\mathrm{and}\ \qR = \left(
        \begin{array}{cc}
          R_{11} & \mb{l}_2^\dagger \\
          \mb{l}_2 & \qR_1 \\
        \end{array}
      \right).
}
Then $b_1 = \widetilde{\qw}^\dagger\mb{l}_1$, $c_1=\widetilde{\qw}^\dagger\qQ_1\widetilde{\qw}$, $b_2 = \widetilde{\qw}^\dagger\mb{l}_2$
and $c_2=1+\widetilde{\qw}^\dagger\qR_1\widetilde{\qw}$.


For the solution of (\ref{indSNRmaxAlterMin}) we give the following theorem, the proof of which can be found  in Appendix \ref{Proof:TheoAlterMin}.
\medskip
\begin{theorem}\label{Theo:AlterMin}
{\em If $a_1/a_2 = b_1/b_2 = c_1/c_2$, the objective in (\ref{indSNRmaxAlterMin}) is a constant, and  the optimum $y$, i.e., $y^\star$,
 is any value satisfying $|y|\le \beta$. Otherwise:
If the equation $(a_1-t a_2)\beta^2 + 2|b_1-t b_2|\beta + c_1 - t c_2 = 0$ has a real root, i.e., $t_1$, such that
$|b_1-t_1 b_2| \ge (t_1 a_2 - a_1)\beta$, then the optimal $y$ is given by
\equ{
y^\star = \beta e^{-\mathrm{i}\theta_1}
}
where $\theta_1 \in (-\pi, \pi]$ is the argument of $b_1-t_1 b_2$; Else, let $t_2$ be the root of
$|b_1-t b_2|^2 = (a_1 - t a_2)(c_1 - t c_2)$ such that
$|b_1-t_2 b_2| < (t_2 a_2 - a_1)\beta$, then the optimal $y$ is given by
\equ{
y^\star = \frac{|b_1-t_2 b_2|}{t_2 a_2 - a_1} e^{-\mathrm{i}\theta_2}
}
where $\theta_2$ is the argument of $b_1-t_2 b_2$.
}
\end{theorem}
{\em Remarks}: The roots $t_1$ and $t_2$ in Theorem \ref{Theo:AlterMin} can both be obtained in closed form.
\medskip

For the coordinate descent method, obviously the function value sequence converges.
However, in general additional conditions for convergence to a stationary point (or fixed point used in \cite[\S6]{Leeuw}) are needed.
\medskip

\begin{theorem}\label{GlobConverg}
{\em
The sequence $\{\qw^{k}\}$ generated by Algorithm \ref{algo:CDM} converges globally to a stationary point.
}
\end{theorem}
\medskip

\begin{proof}
Our proof is based on \cite[Proposition 2.7.1]{Bertsekas} and its proof.
Let us denote the objective in (\ref{indSNRmax1}) as $f(\qw)$.
Let $\bar{\qw}=(\bar{w}_1,\cdots,\bar{w}_N)$ be the limit point of the sequence $\{\qw^{k}\}$.
We first show
\equ{
f(\bar{\qw}) \ge f(w_1, \bar{w}_2, \cdots, \bar{w}_N), \ \forall w_1.\label{sta}
}
If $f(w_1, \bar{w}_2, \cdots, \bar{w}_N)$ is a constant, then obviously (\ref{sta}) holds.
If $f(w_1, \bar{w}_2, \cdots, \bar{w}_N)$ is not a constant, to see why,
let us assume that (\ref{sta}) does not hold.
 A verbatim repetition of the proof for \cite[Proposition 2.7.1]{Bertsekas}
results in
\equ{
f(\bar{\qw}) = f(\bar{w}_1 + \epsilon v_1, \bar{w}_2, \cdots, \bar{w}_N), \ \forall\epsilon\in [0,\epsilon_0]\label{Otherwise}
}
for some $v_1\ne 0$, $\epsilon_0>0$.
But from Theorem \ref{Theo:AlterMin}, (\ref{Otherwise}) does not hold for any $v_1\ne 0$, $\epsilon_0>0$
if $f(w_1, \bar{w}_2, \cdots, \bar{w}_N)$ is not a constant. Thus, (\ref{sta}) holds.
Similarly, we show
\equ{
f(\bar{\qw}) \ge f(\bar{w}_1, \cdots, \bar{w}_{j-1}, w_j, \bar{w}_{j+1}, \cdots, \bar{w}_N), \ \forall w_j\label{sta2}
}
for $j=1,\cdots, N$.
This completes the proof.
\end{proof}
\bigskip


\subsubsection{$p$-norm approximation}\label{Sec:p-norm}

If the solution $\qX^\star$ from the CVX software has  rank greater than one,
we can also use $p$-norm approximation plus an augmented Lagrangian method to solve the problem of (\ref{indNotDiagSNRmax2f}).
The convergence of the augmented Lagrangian method can be found in \cite{Nocedal}.
First, we can show that the problem of (\ref{indNotDiagSNRmax2f}) is equivalent ({\em up to scaling}) to
\al{
&\min_{\qw}\ \left[\max_{k\in I}\, \qw^\dagger\qA_k\qw \right]\label{indNotDiagSNRmaxInfNorm2}\\
&\mathrm{s.t.}\quad  \qw^\dagger\qR\qw=1.\nnb
}
To see why this is the case, let $\qw^\star$ be the solution to the problem of (\ref{indNotDiagSNRmaxInfNorm2}) associated with
the optimal objective value $\max_{k\in I}\ {\qw^\star}^\dagger\qA_k\qw^\star=C$.
Then, $\frac{1}{\sqrt{C}}\qw^\star$ is the solution to the problem of (\ref{indNotDiagSNRmax2f}) associated with
optimal objective value $(\frac{1}{\sqrt{C}}\qw^\star)^\dagger\qR(\frac{1}{\sqrt{C}}\qw^\star)=\frac{1}{C}$. Otherwise, let us assume that
the solution to the problem of (\ref{indNotDiagSNRmax2f}) is $\qw'$ with ${\qw'}^\dagger\qR\qw' = C_1 >\frac{1}{C}$.
Thus, $\qw''=\frac{1}{\sqrt{C_1}}\qw'$ satisfies
 ${\qw'}^\dagger\qR\qw'=1$ and $\max_{k\in I}\ {\qw''}^\dagger\qA_k\qw''=\frac{1}{C_1}<C$.
This contradicts the optimality of $\qw^\star$ for the problem of (\ref{indNotDiagSNRmaxInfNorm2}).
In fact, the two problems are equivalent {\em up to scaling}.

On denoting $\qD_1=\mathrm{diag}(\sqrt{\frac{P_s D_{11}+\sigma^2}{P_1}}, \cdots, \sqrt{\frac{P_s D_{NN}+\sigma^2}{P_N}}\,)$,
$\qu=\qD_1\qw$, $\qR_1=\qD_1^{-1}\qR\qD_1^{-1}$ and $\qQ_1=\qD_1^{-1}\qQ\qD_1^{-1}$,
we rewrite the problem of (\ref{indNotDiagSNRmaxInfNorm2}) as
\al{
&\min_{\qu}\ \qu^\dagger\qQ_1\qu + \|\qu\|_{\infty}^2 \label{indNotDiagSNRmaxInfNorm3}\\
&\mathrm{s.t.}\quad  \qu^\dagger\qR_1\qu=1\nnb
}
where $\|\qu\|_{\infty}=\max_{k\in I}\, |u_k|$ is the infinity norm.
Note that $\|\qu\|_{\infty}$ is not smooth \cite{Charalambous2}.
However, we can approximate $\|\qu\|_{\infty}$ by (smooth) $p$-norm, i.e.,
$\|\qu\|_p=(\sum\nolimits_{k\in I} |u_k|^p)^{1/p}$, so that \cite{Cheney}, \cite{Chen}
\al{
&\|\qu\|_{\infty} = \lim_{p\to \infty} \|\qu\|_p,\label{p-normProp2}\\
\mathrm{and}\ &\|\qu\|_{\infty}\le \|\qu\|_p\le N^{1/p}\|\qu\|_{\infty}.\label{p-normProp2}
}
When $p$ is sufficiently large, the approximation is good.
In fact, from (\ref{p-normProp2}), it is easy to show that given a tolerance $\varepsilon$, the relative error does not exceed $\varepsilon$ as long as $p\ge \log N/\log(1+\varepsilon)$. For example, for $N=10$, $\varepsilon=1\%$, we get $p\ge 232$;
for $N=40$, $\varepsilon=0.5\%$, we get $p\ge 740$.

Now, using $\|\qu\|_{2p}^2$, $p\ge 1$ as a smooth approximation to $\|\qu\|_{\infty}^2$,
we turn to solve the following
\al{
&\min_{\qu}\ \qu^\dagger\qQ_1\qu+\|\qu\|_{2p}^2\label{indNotDiagSNRmaxInfNorm4}\\
&\mathrm{s.t.}\quad  \qu^\dagger\qR_1\qu=1.\nnb
}

We use the augmented Lagrangian method \cite[\S17]{Nocedal} to solve the problem of (\ref{indNotDiagSNRmaxInfNorm4}).
Since the augmented Lagrangian method was originally proposed for real variables,
 we first modify our problem as follows. Define \cite{Telatar}
\al{
\qz &= \left(
         \begin{array}{c}
           \mathrm{Re}(\qu) \\
           \mathrm{Im}(\qu) \\
         \end{array}
       \right),\\
\qF &= \left(
          \begin{array}{cc}
            \mathrm{Re}(\qQ_1) & -\mathrm{Im}(\qQ_1) \\
            \mathrm{Im}(\qQ_1) & \mathrm{Re}(\qQ_1) \\
          \end{array}
        \right), \\
\qK &= \left(
          \begin{array}{cc}
            \mathrm{Re}(\qR_1) & -\mathrm{Im}(\qR_1) \\
            \mathrm{Im}(\qR_1) & \mathrm{Re}(\qR_1) \\
          \end{array}
        \right),\\
\mathrm{and}\ \widetilde{\qJ}_k &= \left(
          \begin{array}{cc}
            \qJ_k & 0 \\
            0 & \qJ_k \\
          \end{array}
        \right)
}
where $\qJ_k$ is defined in Lemma \ref{Lem:QCQP}, and
$\mathrm{Re}(\cdot)$, $\mathrm{Im}(\cdot)$ denote the real and imaginary part respectively, then
\al{
\qu^\dagger\qQ_1\qu&=\qz^T\qF\qz, \\
\qu^\dagger\qR_1\qu&=\qz^T\qK\qz,\\
\mathrm{and}\ \|\qu\|_{2p}^2 &= \bigg(\sum_{k\in I} (\qz^T\widetilde{\qJ}_k\qz)^p\bigg)^{1/p}.\label{phi_p}
}
With these,
we rewrite the problem of (\ref{indNotDiagSNRmaxInfNorm4}) as
\al{
&\min_{\qz}\ \qz^T \qF\qz+\phi_p(\qz)\label{indNotDiagSNRmaxInfNorm5}\\
&\mathrm{s.t.}\quad  \qz^T\qK\qz - 1 = 0\nnb
}
where $\phi_p(\qz)$ is defined as the right hand side of (\ref{phi_p}).

Now we can apply the augmented Lagrangian method, given by
\equ{
L(\qz; \lambda; \mu) = \qz^T \qF\qz+\phi_p(\qz) - \lambda (\qz^T\qK\qz - 1) + \frac{1}{2\mu}(\qz^T\qK\qz - 1)^2\label{augmentedLagrangian}
}
where $\lambda$ is the Lagrangian multiplier,  and the fourth term in the right hand side of (\ref{augmentedLagrangian})
is the penalty function.
The algorithm is described as follows:
{\em
\begin{enumerate}
  \item [1)] Choose an initial estimate $\lambda^{(0)}$ of $\lambda^\star$ and $\mu = 0.001$. Set $k=1$.
  \item [2)] Determine $\qz_k$ to be a minimizer of $L(\qz; \lambda^{(k-1)}; \mu)$;
  \item [3)] Compute $\lambda^{(k)} = \lambda^{(k-1)} - (\qz_k^T\qK\qz_k - 1)/\mu$;
  \item [4)] If a convergence test is satisfied, stop;
  \item [5)] $k=k+1$; Go to 2).
\end{enumerate}
}
For Step 2, we use the backtracking line search Newton's method with Hessian modification \cite[Algorithm 3.2]{Nocedal}.
The iteration expression is
\al{
\qz^{(i+1)}&=\qz^{(i)} + \alpha\mb{p}_i\\
\mathrm{with}\ \mb{p}_i &= -(\nabla^2 L + \beta\qI)^{-1}\nabla L
}
where $\beta$ is chosen such that $\nabla^2 L + \beta\qI$ is positive definite,
e.g., $\beta = \lambda_{\min}(\nabla^2 L)+10^{-6}$, and $\alpha$ is the step size
determined by the backtracking line search described as follows \cite[Algorithm 3.1]{Nocedal}:
{\em
\begin{enumerate}
  \item [a)] Set $\alpha=1$, $c_1=10^{-4}$, $\rho = 0.5$;
  \item [b)] Repeat: if
  $L(\qz^{(i)}+\alpha\mb{p}_i; \cdot; \cdot) > L(\qz^{(i)}; \cdot; \cdot)+ c_1\alpha\mb{p}_i^T\nabla L$,
  then $\alpha \leftarrow \rho\alpha$.
\end{enumerate}
}

In the algorithm, we need  to calculate $\nabla L$ and $\nabla^2 L$ given by
\al{
\nabla L &= 2\qF\qz+\nabla \phi_p - 2\lambda \qK\qz + \frac{2}{\mu}(\qz^T\qK\qz - 1)\qK\qz \\
\mathrm{and}\ \nabla^2 L & = 2\qF + \nabla^2 \phi_p - 2\lambda\qK + \frac{2}{\mu}(\qz^T\qK\qz - 1)\qK
           + \frac{4}{\mu}\qK\qz\qz^T\qK.
}
The calculation of $\nabla \phi_p$ and $\nabla^2 \phi_p$ is given in Appendix \ref{CalcDfD2f}.\\
{\em Remarks}: For the initial estimate $\lambda^{(0)}$ of $\lambda^\star$, note that when $p=1$,
then $\phi_p(\qz) = \qz^T\qz$ and
 $\lambda^\star$ can be expressed in closed form as $\lambda_{\min}(\qK^{-1/2}\qF\qK^{-1/2}+\qK^{-1})$.
We choose this as $\lambda^{(0)}$.

\section{Numerical Results}\label{Sec:Sim}

In this section, we provide some examples  illustrating the proposed algorithms.
For more simulation results  on beamforming itself the reader can refer to \cite{Luo}.
We consider a channel model as follows:
\al{
f_i &= \bar{f}_i + \sqrt{\psi_i}\,\tilde{f}_i\\
\mathrm{and}\ g_j &= \bar{g}_j + \sqrt{\varphi_j}\,\tilde{g}_j
}
where $\bar{f}_i$ and $\bar{g}_j$ are means, $\psi_i$ and $\varphi_j$ are variances,
$\tilde{f}_i$ and $\tilde{g}_j$ both are zero-mean
random variables with unit variance.
We assume that $\tilde{f}_i$, $\tilde{f}_j$, $\tilde{g}_i$ and $\tilde{g}_j$, $\forall i\ne j$ are independent.
$\bar{f}_i=0$ corresponds to the scenario in which there is no line-of-sight (LOS) path (Rayleigh fading),
while $\bar{f}_i\ne 0$ corresponds the scenario in which there is an LOS path (Rician fading).
Thus, the matrices $\qD$, $\qR$ and $\qQ$ are given by
\al{
\qD &= \mathrm{diag}(|\bar{f}_1|^2+\psi_1, \cdots, |\bar{f}_N|^2+\psi_N),\nnb\\
Q_{ij} &= \bar{g}_i\bar{g}_j^\ast + \sqrt{\varphi_i\varphi_j}\,\delta_{ij},\nnb\\
\mathrm{and}\ R_{ij} &= (\bar{f}_i\bar{f}_j^\ast + \sqrt{\psi_i\psi_j}\,\delta_{ij})(\bar{g}_i\bar{g}_j^\ast
 + \sqrt{\varphi_i\varphi_j}\,\delta_{ij})\nnb
}
where $\delta_{ij}$ is the Kronecker function.


\subsection{SNR maximization under total power constraint}

Please refer to \S\ref{Sec:SNRmaxTotalPower} for details.
First, we consider a network consisting of $N=6$ relays with channel parameters given by
\medskip

\begin{figure}[hbtp]
\centering
\includegraphics[width=3.5in]{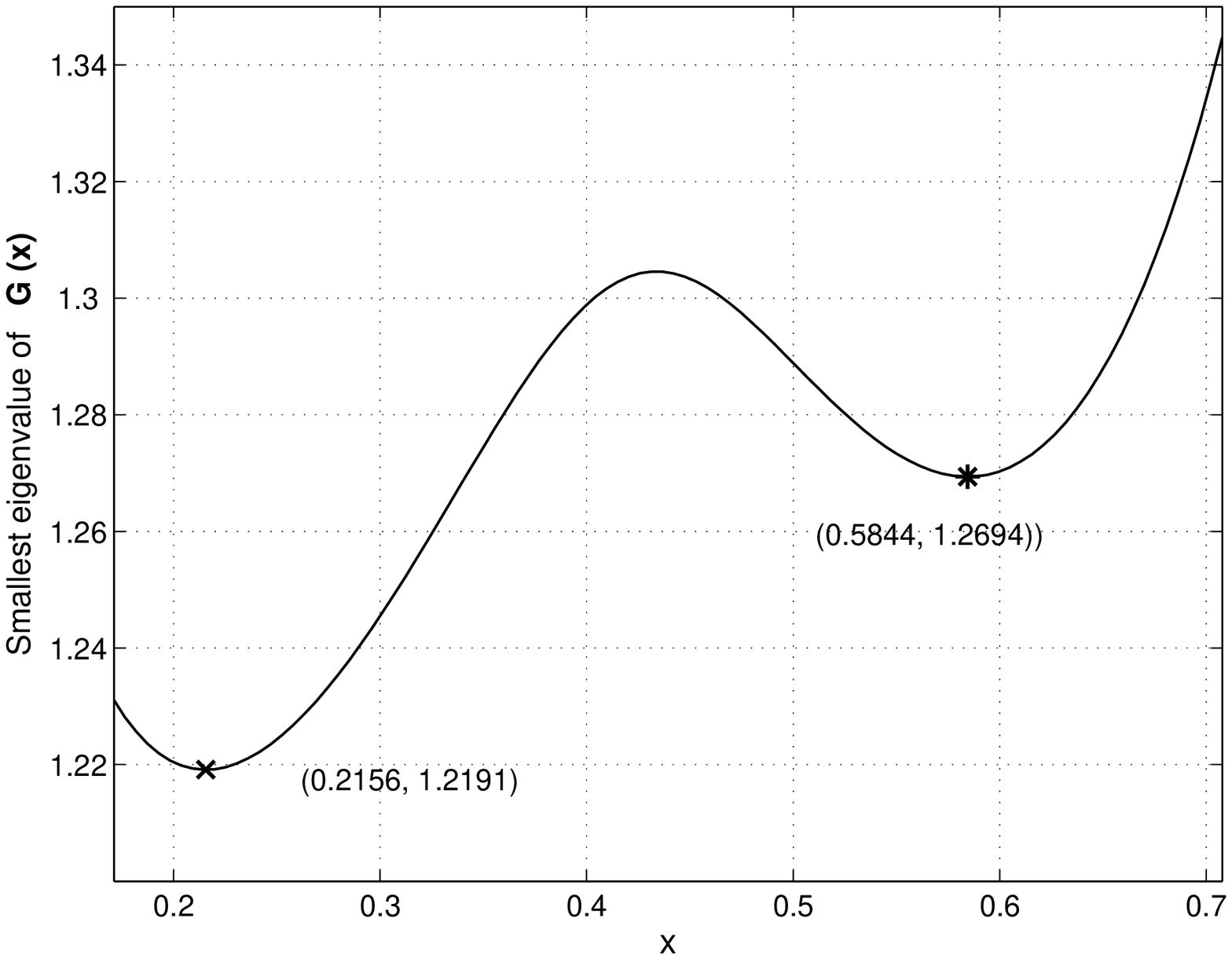}
\caption{$\lambda_{\min}(\qG(x))$ for $x$ in $[x_l, x_u]=[0.2754, 0.6392]$ with $100$ uniform points;
The left point is for starting point $x_0=x_l$ and the right point is
for the starting point $x_0=x_u$; $\mathrm{SNR}=10\,\mathrm{dB}$;  A total power constraint.}
\label{fig:2}
\end{figure}

\begin{figure}[hbtp]
\centering
\includegraphics[width=3.5in]{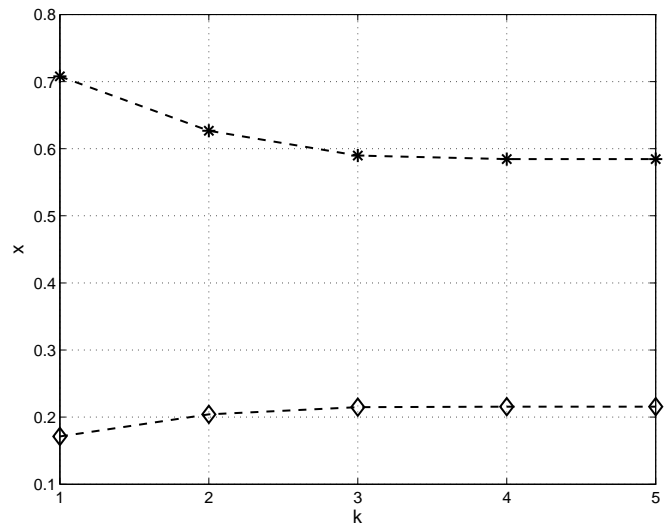}
\caption{The iteration process; The upper line is for starting point $x_0=x_l$ and the lower line is
for the starting point $x_0=x_u$; $\mathrm{SNR}=10\,\mathrm{dB}$;  A total power constraint.}
\label{fig:3}
\end{figure}

{{
$\bar{f}_1 = \ \ 0.2202 + 0.8130\mathrm{i}, \ \bar{f}_2 = -0.4075 - 0.7644\mathrm{i},$

$\bar{f}_3 = -2.0107 + 0.4016\mathrm{i}, \ \bar{f}_4 = -0.4503 + 0.0678\mathrm{i}, $

$\bar{f}_5 = \ \ 0.8588 - 0.1130\mathrm{i}, \ \bar{f}_6 = -0.1219 + 0.4260\mathrm{i};$
\medskip

$\psi_1 = 3.8042, \ \psi_2 = 2.6326, \ \psi_3 = 4.7590, $

$\psi_4 = 0.4989, \ \psi_5 = 1.2576, \ \psi_6 = 1.2484;$

\medskip

$\bar{g}_1 = -0.3726 + 0.8007\mathrm{i}, \ \bar{g}_2 = \ \ 0.4592 - 0.2045\mathrm{i},$

$\bar{g}_3 = -0.8769 + 0.4671\mathrm{i}, \ \bar{g}_4 = -0.9270 + 0.5430\mathrm{i}, $

$\bar{g}_5 = -0.0063 - 0.4977\mathrm{i}, \ \bar{g}_6 = -0.7783 - 0.7712\mathrm{i};$

\medskip

$\varphi_1 = 0.3913, \ \varphi_2 = 0.4791, \ \varphi_3 = 0.0865, $

$\varphi_4 = 2.7813, \ \varphi_5 = 4.8960, \ \varphi_6 = 4.6789$.
}
\medskip\\
}
Fig. \ref{fig:2} plots $\lambda_{\min}(\qG(x))$ for $x$ in $[x_l, x_u]=[0.1711, 0.7077]$ with $100$ uniform points.
Using Newton's method for starting points $x_0=x_l$, $x_0=x_u$,
the convergent points $(0.2156, 1.2191)$, $(0.5844, 1.2694)$ are also plotted in Fig. \ref{fig:2}.
Fig. \ref{fig:3} plots the iteration process under the stopping test: $|\frac{x_{k+1}-x_k}{x_k}| < 10^{-3}$
and $|\frac{\mathrm{d}}{\mathrm{d} x}\lambda_{\min}(\qG(x))|<10^{-3}$.
It can be seen from Fig. \ref{fig:3} that Newton's method converges rapidly.

Second, we consider a network consisting of $N=6$ relays with channel parameters given by
\medskip

\begin{figure}[hbtp]
\centering
\includegraphics[width=3.5in]{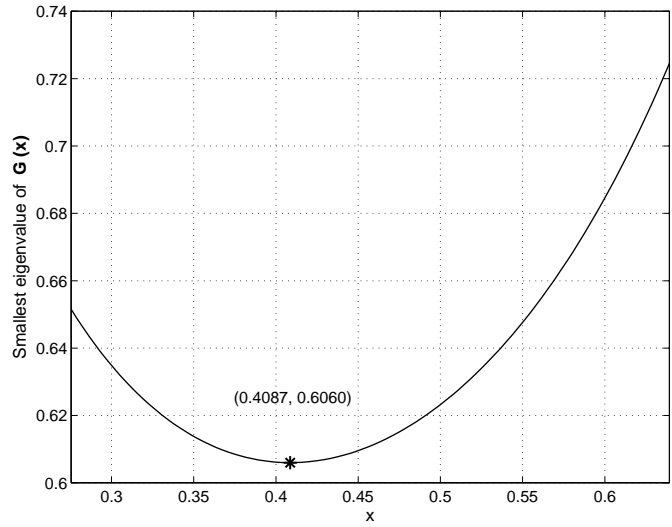}
\caption{$\lambda_{\min}(\qG(x))$ for $x$ in $[x_l, x_u]=[0.2754, 0.6392]$ with $100$ uniform points;
The same convergent point is for starting point $x_0=x_l$ and $x_0=x_u$; $\mathrm{SNR}=10\,\mathrm{dB}$;
A total power constraint.}
\label{fig:4}
\end{figure}

\begin{figure}[hbtp]
\centering
\includegraphics[width=3.5in]{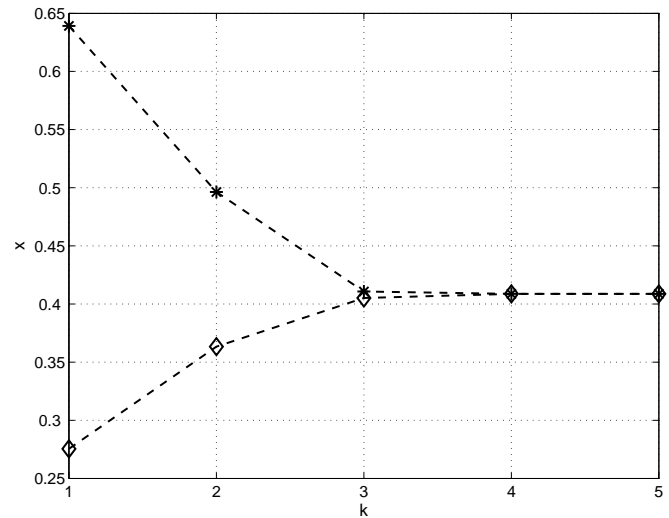}
\caption{The iteration process; The upper and lower lines converge to the same point for starting point $x_0=x_l$ and
$x_0=x_u$; $\mathrm{SNR}=10\,\mathrm{dB}$; A total power constraint.}
\label{fig:5}
\end{figure}

{{
$\bar{f}_1 = -0.4751 + 0.7340\mathrm{i}, \ \bar{f}_2 = -0.0449 - 0.4609\mathrm{i},$

$\bar{f}_3 = \ \ 0.0239 - 1.5154\mathrm{i}, \ \bar{f}_4 = \ \ 0.5130 - 0.1755\mathrm{i}, $

$\bar{f}_5 = -0.2017 + 0.6717\mathrm{i}, \ \bar{f}_6 = \ \ 1.0134 - 0.1985\mathrm{i};$
\medskip

$\psi_1 = 2.4707, \ \psi_2 = 3.9193, \ \psi_3 = 2.4121, $

$\psi_4 = 3.8879, \ \psi_5 = 1.2050, \ \psi_6 = 3.0901;$

\medskip

$\bar{g}_1 = \ \ 0.5360 - 1.2932\mathrm{i}, \ \bar{g}_2 = \ \ 1.7471 - 0.8914\mathrm{i},$

$\bar{g}_3 = \ \ 0.0955 - 0.1577\mathrm{i}, \ \bar{g}_4 = -0.6795 + 0.2479\mathrm{i}, $

$\bar{g}_5 = \ \ 0.5815 + 0.5039\mathrm{i}, \ \bar{g}_6 = -0.3090 + 0.8413\mathrm{i};$

\medskip

$\varphi_1 = 3.9655, \ \varphi_2 = 0.2693, \ \varphi_3 = 0.9205, $

$\varphi_4 = 0.5567, \ \varphi_5 = 3.3901, \ \varphi_6 = 2.9367$.
}
\medskip\\
}
Fig. \ref{fig:4} plots $\lambda_{\min}(\qG(x))$ for $x$ in $[x_l, x_u]=[0.2754, 0.6392]$ with $100$ uniform points.
Using Newton's method for starting points $x_0=x_l$, $x_0=x_u$,
the same convergent point $(0.4087, 0.6060)$ is also plotted in Fig. \ref{fig:4}.
Fig. \ref{fig:5} plots the iteration process under the stopping test: $|\frac{x_{k+1}-x_k}{x_k}| < 10^{-3}$
and $|\frac{\mathrm{d}}{\mathrm{d} x}\lambda_{\min}(\qG(x))|<10^{-3}$.
It can be seen from Fig. \ref{fig:5} that Newton's method converges rapidly.

\subsection{SNR maximization under individual relay power constraints}

Please refer to \S\ref{Sec:SNRmaxIndiRelayPower} for details.
In \cite{Luo}, the authors stated that, based on their simulations, the SDP relaxation always has a rank one solution. However, no analytic proof was provided  for that claim.
However, although a rank one solution often occurs, for general $\qR$ and $\qQ$ the SDP relaxation does not necessarily have a rank one solution.
This can be seen in the following examples,
 for which the SDP relaxation has a rank greater than one.

First, we consider a network consisting of $N=4$ relays with
{\small\al{
\qQ &=
         \left(\begin{array}{rrrr}
           2.1 & .73+.75\mathrm{i} & .43+1.1\mathrm{i} & .70-.33\mathrm{i} \\
           .73-.75\mathrm{i} & 1.6 & -.20+.18\mathrm{i} & .57-.71\mathrm{i} \\
           .43-1.1\mathrm{i} & -.20-.18\mathrm{i} & 2 & -.52-.45\mathrm{i} \\
           .70+.33\mathrm{i} & .57+.71\mathrm{i} & -.52+.45\mathrm{i} & .98 \\
         \end{array}\right)
\\
\mathrm{and}\ \qR &=
         \left(\begin{array}{rrrr}
           1.6 & -.74-.16\mathrm{i} & .084-.57\mathrm{i} & -.19+.67\mathrm{i} \\
           -.74+.16\mathrm{i} & 1.1 & -.88+.31\mathrm{i} & -.44-.24\mathrm{i} \\
           .084+.57\mathrm{i} & -.88-.31\mathrm{i} & 2 & .20-.14\mathrm{i} \\
           -.19-.67\mathrm{i} & -.44+.24\mathrm{i} & .20+.14\mathrm{i} & 1.5\\
         \end{array}\right).
}
}

\begin{figure}[hbtp]
\centering
\includegraphics[width=3.5in]{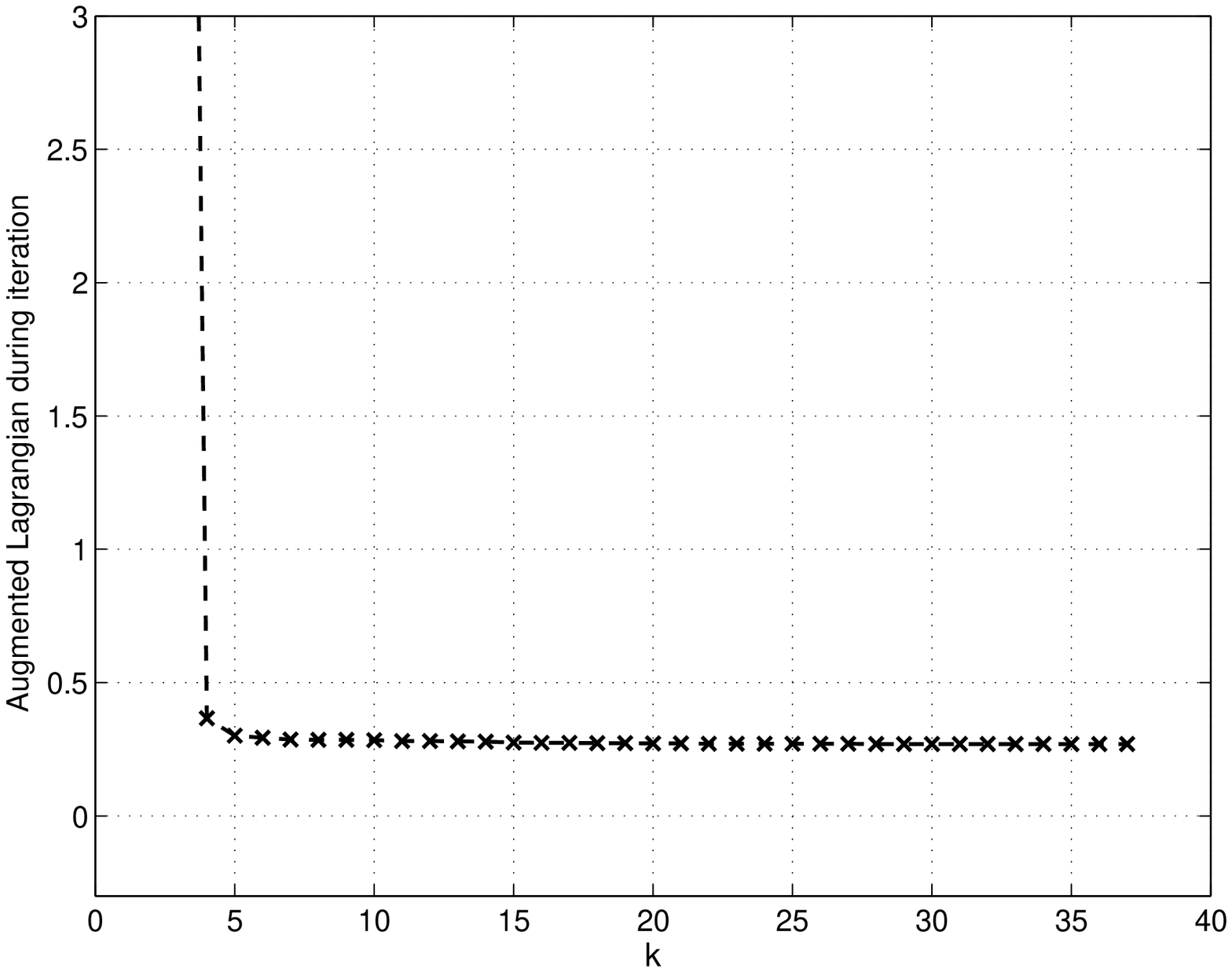}
\caption{The augmented Lagrangian $L(\qz; \lambda; \mu)$ (defined in (\ref{augmentedLagrangian}))
during the iteration process of the proposed algorithm; Individual relay power constraints; $p$-norm approximation.}
\label{fig:6}
\end{figure}

\begin{figure}[hbtp]
\centering
\includegraphics[width=3.5in]{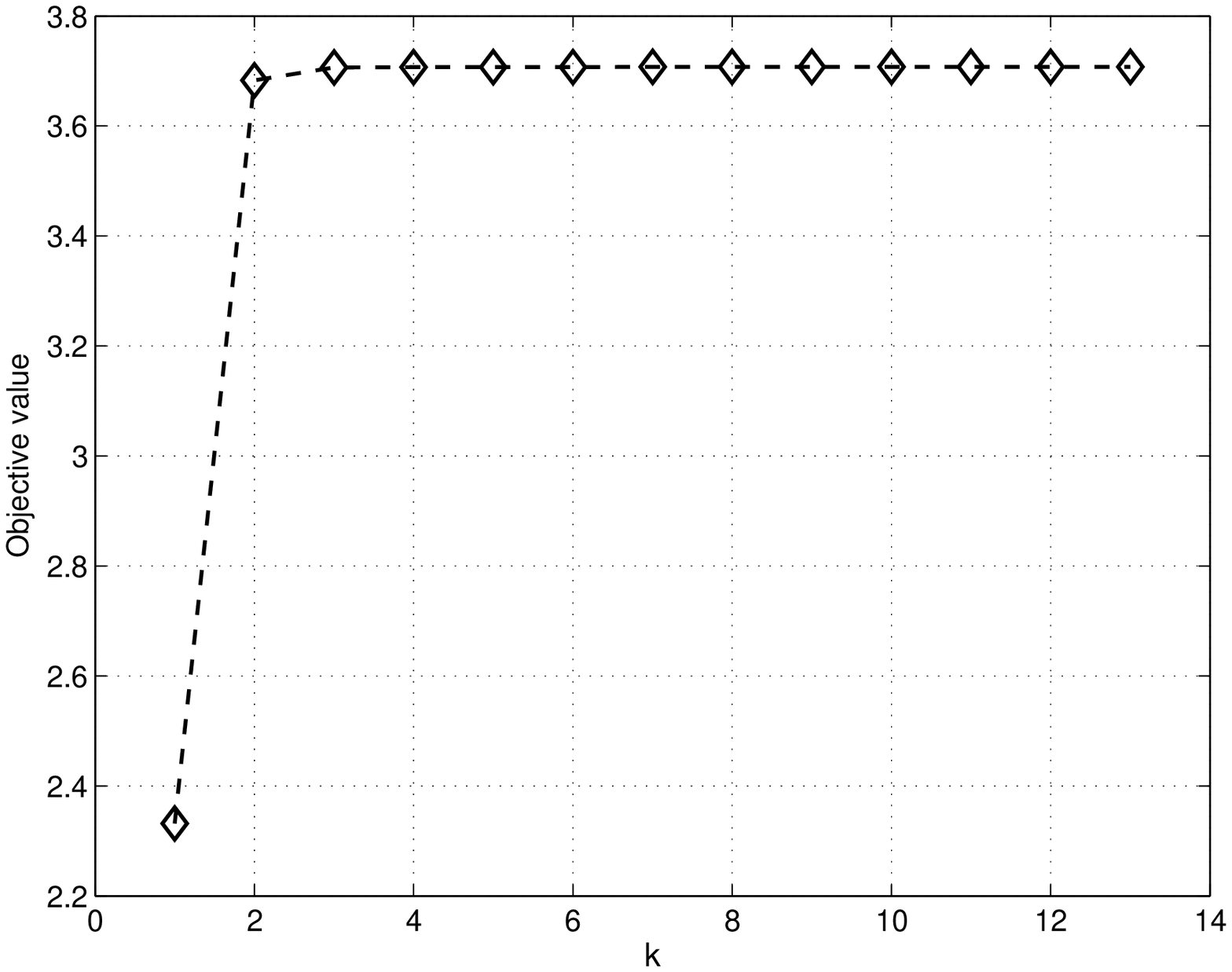}
\caption{The objective value
during the iteration process of the proposed algorithm; Individual relay power constraints; Coordinate descent method.}
\label{fig:7}
\end{figure}

For simplicity, we let $\qD_1=\qI$ (defined in \S\ref{Sec:p-norm}) and
denote the SDP relaxation solution from CVX software by $\qX^\star$.
The eigenvalues of $\qX^\star$ are
\equ{
0.0000, 0.0000, 0.2064, 1.8148.\nnb
}
Thus, $\qX^\star$ has rank two rather than rank one and can be eigen-decomposed as
$0.2064\qu_1\qu_1^\dagger+ 1.8148\qu_2\qu_2^\dagger$ where $\qu_1$ and $\qu_2$ are eigenvectors
associated with the eigenvalues $0.2064$ and $1.8148$ respectively.
We obtain the objective values of the problem of (\ref{indNotDiagSNRmax2f}) for SDP relaxation,
GRP from \cite{Luo},
coordinate descent method from \S\ref{Sec:AlterativeMin},
and $p$-norm approximation from \S\ref{Sec:p-norm} (starting points: $\sqrt{1.8148}\,\qu_2$ or some samples from $\mathcal{CN}(0, \qX^\star)$)
as, respectively:

\medskip
{\em SDP relaxation: \ $3.74112$

GRP ($10^6$ samples from $\mathcal{CN}(0, \qX^\star)$): \ $3.6970$

Coordinate descent method: \ $3.7076$

$p$-norm approximation: \ $3.7069$ ($p=1024$)}
\medskip\\
It can be seen that the objective values from GRP, coordinate descent method and $p$-norm approximation are close to each other (with a difference $<0.3\%$)
and close to the SDP relaxation solution (with a difference $<2\%$).
It can be seen that: although the GRP attains a close performance compared with the other two methods, it is
time consuming in the sense that it needs much more time (processes $10^6$ samples from $\mathcal{CN}(0, \qX^\star)$).
The augmented Lagrangian $L(\qz; \lambda; \mu)$ (defined in (\ref{augmentedLagrangian})) during the iteration is plotted in Fig. \ref{fig:6}. The objective value during the iteration for the coordinate descent method is plotted in Fig. \ref{fig:7}. It can be seen that
for these two algorithms the iteration converges rapidly.

Second, we consider a network consisting of $N=6$ relays with
{\small
\al{
\qQ &= \left(
         \begin{array}{rrrrrr}
.778&   -.658-.646\mathrm{i}&  .135+.269\mathrm{i}&    -.273+.005\mathrm{i}&     .088-.261\mathrm{i}& -.021-.013\mathrm{i}\\
-.658+.646\mathrm{i}&      2.20&    -.379-1.14\mathrm{i}&     .253-.872\mathrm{i}&       -.337+1.02\mathrm{i}&     .444-.035\mathrm{i}\\
.135-.269\mathrm{i}&       -.379+1.14\mathrm{i}&     2.&        .689+.298\mathrm{i}&       -.547-.160\mathrm{i}&        .373+.693\mathrm{i}\\
-.273-.005\mathrm{i}&        .253+.872\mathrm{i}&        .689-.298\mathrm{i}&      1.&       -.655+.192\mathrm{i}&        .132-.107\mathrm{i}\\
.088+.261\mathrm{i}&       -.337-1.02\mathrm{i}&       -.547+.160\mathrm{i}&       -.655-.192\mathrm{i}&    2.40&       -.721-.276\mathrm{i}\\
-.021+.013\mathrm{i}&     .444+.035\mathrm{i}&        .373-.693\mathrm{i}&   .132+.107\mathrm{i}&   -.721+.276\mathrm{i}&    1.09\\
         \end{array}
       \right)\\
\mathrm{and}\nnb\\
\qR & = \left(
          \begin{array}{rrrrrr}
3.44& -.263+.054\mathrm{i}&     .572+1.73\mathrm{i}&     .490-.276\mathrm{i}&    -.613-1.62\mathrm{i}& -.014+.375\mathrm{i}\\
-.263-.054\mathrm{i}&            3.09&    -.342-1.49\mathrm{i}&     .926+1.13\mathrm{i}&    -.282-.713\mathrm{i}&    -.211+.911\mathrm{i}\\
.572-1.73\mathrm{i}&    -.342+1.49\mathrm{i}&            2.70&    -.493+.865\mathrm{i}&    -.396+.826\mathrm{i}&     .149-.836\mathrm{i}\\
.490+.276\mathrm{i}&     .926-1.13\mathrm{i}&    -.493-.865\mathrm{i}&            3.09&     .541+.330\mathrm{i}&    -.552-.221\mathrm{i}\\
-.613+1.62\mathrm{i}&    -.282+.713\mathrm{i}&    -.396-.826\mathrm{i}&     .541-.330\mathrm{i}&            2.75&    -.442-.352\mathrm{i}\\
-.014-.375\mathrm{i}&    -.211-.911\mathrm{i}&     .149+.836\mathrm{i}&    -.552+.221\mathrm{i}&    -.442+.352\mathrm{i}&            2.08
          \end{array}
        \right).
}
}

\begin{figure}[hbtp]
\centering
\includegraphics[width=3.5in]{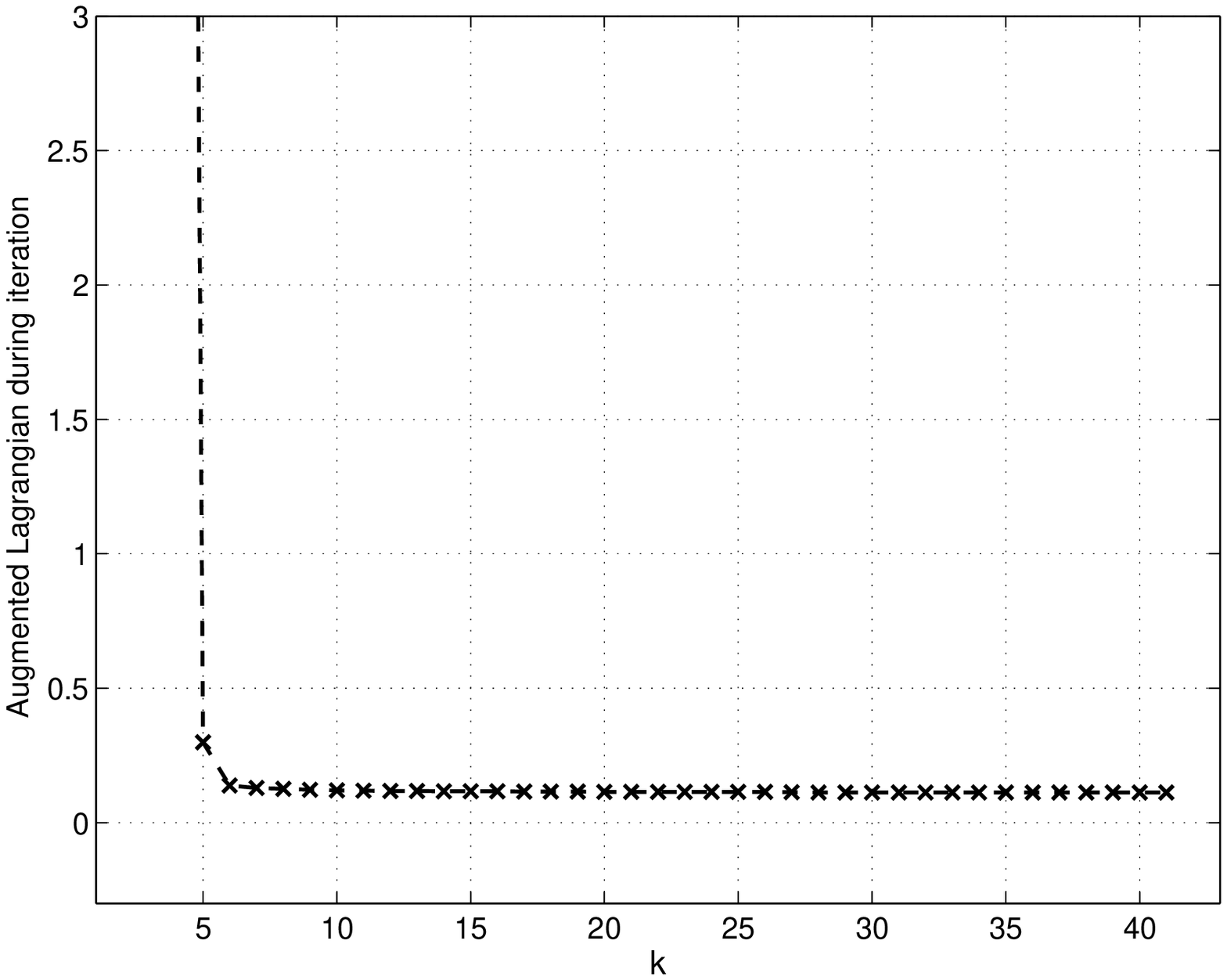}
\caption{The augmented Lagrangian $L(\qz; \lambda; \mu)$ (defined in (\ref{augmentedLagrangian}))
during the iteration process of the proposed algorithm; Individual relay power constraints; $p$-norm approximation.}
\label{fig:8}
\end{figure}

\begin{figure}[hbtp]
\centering
\includegraphics[width=3.5in]{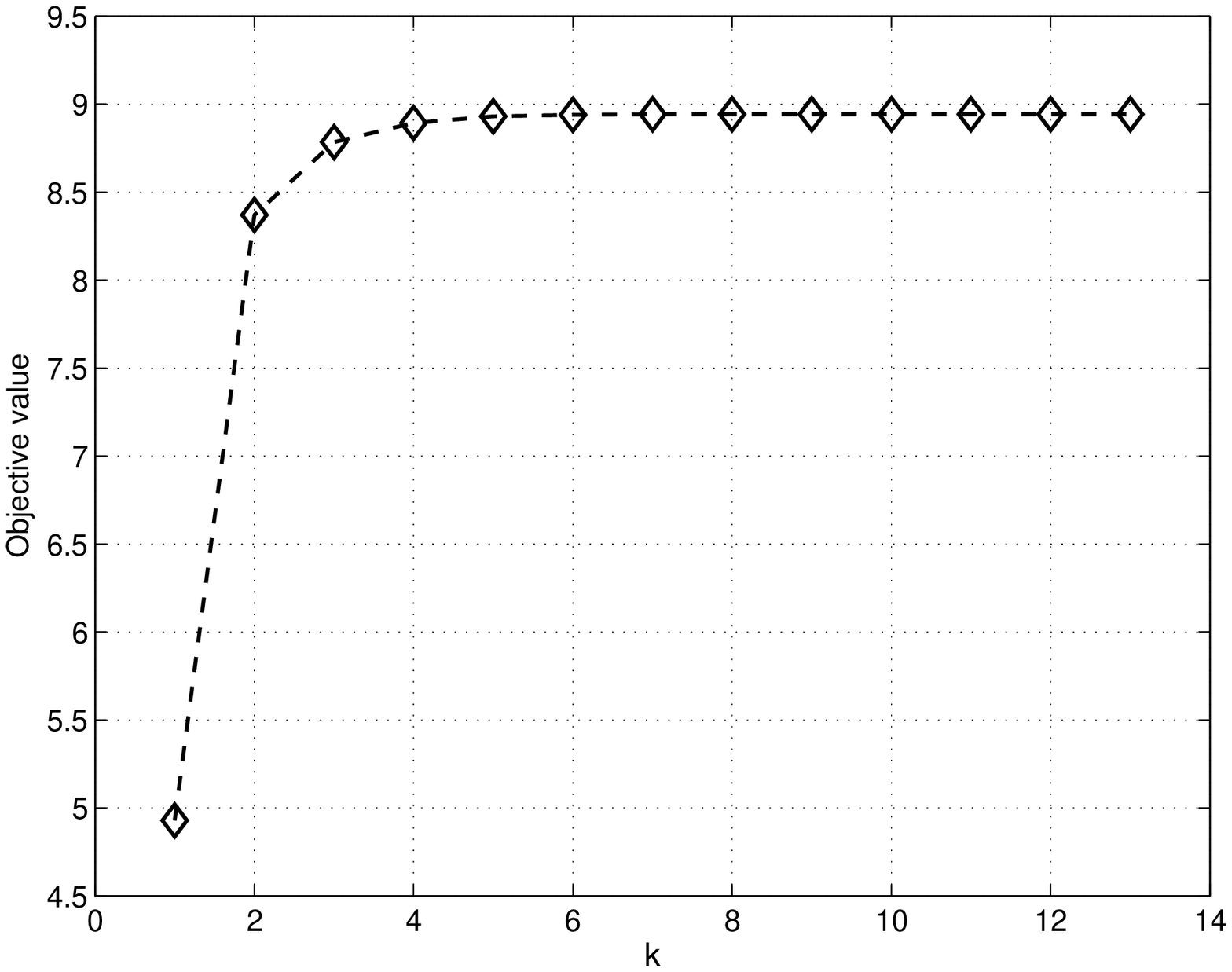}
\caption{The objective value
during the iteration process of the proposed algorithm; Individual relay power constraints; Coordinate descent method.}
\label{fig:9}
\end{figure}

The eigenvalues of $\qX^\star$ are
\equ{
0.0000, 0.0000, 0.0000, 0.0000, 0.8369, 2.3774.\nnb
}
Thus, $\qX^\star$ has rank two rather than rank one and can be eigen-decomposed as
$0.8369\qu_1\qu_1^\dagger+ 2.3774\qu_2\qu_2^\dagger$ where $\qu_1$ and $\qu_2$ are eigenvectors
associated with the eigenvalues $0.8369$ and $2.3774$ respectively.
We obtain the objective values of the problem of (\ref{indNotDiagSNRmax2f}) for SDP relaxation,
GRP, coordinate descent method,
and $p$-norm approximation (starting points: $\sqrt{2.3774}\,\qu_2$
or some samples from $\mathcal{CN}(0, \qX^\star)$) as, respectively:

\medskip
{\em SDP relaxation: \ $9.33816$

GRP ($10^6$ samples from $\mathcal{CN}(0, \qX^\star)$): \ $8.1472$

Coordinate descent method: \ $8.9428$

$p$-norm approximation: \ $8.9409$ ($p=1024$)}
\medskip\\
It can be seen that: the objective value from GRP has a significant ($>10\%$) difference from the SDP relaxation solution;
$p$-norm approximation and coordinate descent method attain objective values close to each other;
the improvement of objective value from coordinate descent method is $(8.9428-8.1472)/8.1472=9.77\%$
compared with GRP.
It can be seen that for this example, GRP is time consuming and ineffective in the sense that it needs more time (processes $10^6$ samples
from $\mathcal{CN}(0, \qX^\star)$) but attains worse performance compared with the other two algorithms.
The augmented Lagrangian $L(\qz; \lambda; \mu)$ (defined in (\ref{augmentedLagrangian})) during the iteration is plotted
in Fig. \ref{fig:8}. The objective value during the iteration for the coordinate descent method is plotted in Fig. \ref{fig:9}.
We can see that for these two algorithms the iteration converges rapidly.

\section{Conclusion}\label{Sec:Conclu}

We  have investigated the problem of cooperative beamforming under the assumption that the second-order statistics
of the channel state information (CSI) are available.
Beamforming weights are determined so that the SNR at the destination is maximized subject to two kinds of power constraints.
The first kind of power constraint is a constraint on the total power, i.e., source plus relay power.
The second kind of power constraint is a constraint on each relay's transmit power.
For uncorrelated Rayleigh fading scenario, we attained the exact solution.
For generic fading scenario, we focused on the case in which the SDP relaxation does not produce a rank-one solution
and proposed two methods to solve it.
The numerical simulations suggest that the proposed methods are more effective than the method of \cite{Luo}.

\appendices

\section{Proof of Lemma \ref{Lem:totSNRmax}}\label{ProofLem:totSNRmax}

Let $(P_s^\circ, \qw^\circ)$ be the solution to the problem of (\ref{totSNRmax}).
We can show that $P_s^\circ+P_s^\circ{\qw^\circ}^\dagger\qD{\qw^\circ} + \sigma^2{\qw^\circ}^\dagger{\qw^\circ} = P_0$.
Otherwise, let us assume that
$P_s^\circ+P_s^\circ{\qw^\circ}^\dagger\qD{\qw^\circ} + \sigma^2{\qw^\circ}^\dagger{\qw^\circ} < P_0$.
Let $\beta=(P_0-P_s^\circ)/(P_s^\circ{\qw^\circ}^\dagger\qD{\qw^\circ} + \sigma^2{\qw^\circ}^\dagger{\qw^\circ})$, and hence $\beta>1$.
It is easy to verify that $(P_s^\circ, \sqrt{\beta}\,\qw^\circ)$ satisfies the constraint but results in a larger objective value.
This violates the optimality of $(P_s^\circ, \qw^\circ)$.
With this, the problem of (\ref{totSNRmax}) is equivalent to
\al{
&\max_{P_s, \qw}\ \frac{P_s}{\sigma^2}\frac{\qw^\dagger\qR\qw}{1+\qw^\dagger\qQ\qw}\label{totSNRmax1}\\
&\mathrm{s.t.} \quad P_s+P_s\qw^\dagger\qD\qw + \sigma^2\qw^\dagger\qw = P_0.\nnb
}
It follows from the constraint in (\ref{totSNRmax1}) that
\equ{
1=\frac{\qw^\dagger(P_s\qD+\sigma^2\qI)\qw}{P_0-P_s}.\label{constraintEq}
}
By using (\ref{constraintEq}),
we rewrite the problem of (\ref{totSNRmax1}) as
\al{
&\max_{P_s, \qw}\ \frac{P_s}{\sigma^2}\frac{(P_0-P_s)\qw^\dagger\qR\qw}{
\qw^\dagger[P_s\qD+\sigma^2\qI+(P_0-P_s)\qQ]\qw}\label{totSNRmax2}\\
&\mathrm{s.t.} \quad P_s+P_s\qw^\dagger\qD\qw + \sigma^2\qw^\dagger\qw = P_0.\nnb
}
Note that the objective in (\ref{totSNRmax2}) has the same value at $\qw$ and $\beta_1\qw$, $\forall \beta_1\ne 0$, $\qw\ne 0$.
Thus, the problem of (\ref{totSNRmax2}) is equivalent to
\al{
&\max_{P_s, \qw}\ \frac{P_s}{\sigma^2}\frac{(P_0-P_s)\qw^\dagger\qR\qw}{
\qw^\dagger[P_s\qD+\sigma^2\qI+(P_0-P_s)\qQ]\qw}\label{totSNRmax3}\\
&\mathrm{s.t.} \quad 0\le P_s\le P_0, \ \qw\ne 0.\nnb
}

Further, we rewrite
\equ{
\sigma^2\qI=P_s\frac{\sigma^2}{P_0}\qI+(P_0-P_s)\frac{\sigma^2}{P_0}\qI,
}
which enables us to write
\equ{
 P_s\qD+\sigma^2\qI+(P_0-P_s)\qQ
=  P_s\left(\qD+\frac{\sigma^2}{P_0}\right)+ (P_0-P_s)\left(\qQ+\frac{\sigma^2}{P_0}\right).
}
With this, by using the fact that for $\qC_1\succ 0$ and $\qC_2\succ 0$ \cite[p. 549]{Meyer}
\equ{
\frac{1}{\lambda_{\min}(\qC_1^{-1/2}\qC_2\qC_1^{-1/2})}=\max_{\qx\ne 0}\ \frac{\qx^\dagger\qC_1\qx}{\qx^\dagger\qC_2\qx},
\label{mineigen}
}
the problem of (\ref{totSNRmax3}) is equivalent to the problem of (\ref{totSNRmax4}).

\section{Proof of Lemma \ref{Lem:Opt_x_interval}}\label{ProofLem:Opt_x_interval}

Obviously, neither $x=0$ nor $x=1$ is the solution to the problem of (\ref{totSNRmax4a}).
Define the function
\equ{
K(x)=\frac{x\qS_1+(1-x)\qS_2}{x(1-x)}=\frac{\qS_1}{1-x}+\frac{\qS_2}{x}, \ x\in (0,1).\label{Kx}
}
Let $x\in (0,1)$ and $\Delta x\ne 0$ be an increment such that $x+\Delta x \in (0,1)$. Using Taylor series expansion, we approximate
\al{
\frac{1}{1-(x+\Delta x)}&=\frac{1}{1-x}+ \frac{\Delta x}{(1-x)^2} + \frac{(\Delta x)^2}{(1-\xi_1)^3},\label{Taylor1}\\
\mathrm{and}\ \frac{1}{x+\Delta x}&= \frac{1}{x} - \frac{\Delta x}{x^2} + \frac{(\Delta x)^2}{\xi_2^3}\label{Taylor2}
}
where $\xi_1$ and $\xi_2$ both lie between $x$ and $x+\Delta x$. From (\ref{Kx}), (\ref{Taylor1}) and (\ref{Taylor2}), we get
\equ{
K(x+\Delta x)=K(x)+\Delta x \left(\frac{1}{(1-x)^2}\qS_1-\frac{1}{x^2}\qS_2\right)
 +(\Delta x)^2\left(\frac{1}{(1-\xi_1)^3}\qS_1+\frac{1}{\xi_2^3}\qS_2\right).\label{TaylorExpan}
}
Note that the third term in the right hand side of (\ref{TaylorExpan}) is positive definite.
By using the facts that \cite[p. 549]{Meyer}
\equ{
c\qS_1\preceq \qS_2, \ \mathrm{and}\ \ d\qS_1\succeq \qS_2
}
it is not difficult to prove that
\equ{
\frac{1}{(1-x)^2}\qS_1-\frac{1}{x^2}\qS_2\ \left\{\begin{array}{cc}
                                                    \preceq 0 & x\in (0, x_l] \\
                                                    \succeq 0 & x\in [x_u, 1).
                                                  \end{array}
\right.
}
With these, we know that: if $x\in (0, x_l]$ and $\Delta x <0$, then $K(x+\Delta x) \succ K(x)$ and it follows from
Weyl's inequality \cite[p. 181]{Horn} that $\lambda_{\min}(K(x+\Delta x))>\lambda_{\min}(K(x))$;
if $x\in [x_u, 1)$ and $\Delta x >0$, similarly, $\lambda_{\min}(K(x+\Delta x))>\lambda_{\min}(K(x))$.
This completes the proof.

\section{Proof of Theorem \ref{Theo:t_star}}\label{ProofTheo:t_star}

If $k_0=1$, then $0< t^\star < \tilde{t}_1$, and
\equ{
\frac{P_s}{\sigma^2}\tilde{r}_k-t^\star \tilde{q}_k >0, \ k=1,\cdots,N.
}
Thus, $F(t)=0$ in (\ref{F_t_2}) leads to
\equ{
-t^\star+\sum_{k=1}^N\frac{\tilde{P}_k}{P_s \tilde{D}_{kk}+\sigma^2}\bigg(\frac{P_s}{\sigma^2}\tilde{r}_k-t^\star\tilde{q}_k\bigg)=0.
}
The desired result can be obtained from the above equation.

If $k_0>1$, then $\tilde{t}_{k_0-1}< t^\star < \tilde{t}_{k_0}$, and
\equ{
\frac{P_s}{\sigma^2}\tilde{r}_k-t^\star\tilde{q}_k\ \left\{\begin{array}{cc}
                                                     >0 & k=k_0, \cdots, N \\
                                                     <0 & k=1,\cdots,k_0-1.
                                                   \end{array}\right.
}
Thus, $F(t)=0$ in (\ref{F_t_2}) leads to
\equ{
-t^\star+\sum_{k=k_0}^N\frac{\tilde{P}_k}{P_s \tilde{D}_{kk}+\sigma^2}\bigg(\frac{P_s}{\sigma^2}\tilde{r}_k-t^\star\tilde{q}_k\bigg)=0.
}
The desired result can be obtained from the above equation.

\section{Proof of Lemma \ref{Lem:QCQP}}\label{ProofLem:QCQP}

Note that the constraints in (\ref{indSNRmax1}) can be rewritten as
\equ{
\frac{P_s D_{kk}+\sigma^2}{P_k}\qw^\dagger\qJ_k\qw\le 1, \ k\in I.
}
Note that the objective in (\ref{indSNRmax1}) has a greater value at $\alpha\qw$
than that at $\qw$, $\forall \alpha> 1$, $\qw\ne 0$.
Thus, there exists $j\in I$ such that $((P_s D_{jj}+\sigma^2)/P_j)\qw^\dagger\qJ_j\qw= 1$, i.e., at least one constraint is active.
With this, the constraints in (\ref{indSNRmax1}) can be rewritten as
\equ{
\max_{k\in I}\ \frac{P_s D_{kk}+\sigma^2}{P_k}\qw^\dagger\qJ_k\qw = 1.\label{constraintMax}
}
By using (\ref{constraintMax}),
we rewrite the problem of (\ref{indSNRmax1}) as
\al{
&\max_{\qw}\ \frac{P_s}{\sigma^2}\frac{\qw^\dagger\qR\qw}{\max_{k\in I}
\qw^\dagger\qA_k\qw}\label{indNotDiagSNRmax1}\\
&\mathrm{s.t.} \quad \max_{k\in I} \frac{P_s D_{kk}+\sigma^2}{P_k}\qw^\dagger\qJ_k\qw = 1.\nnb
}
Note that the objective in (\ref{indNotDiagSNRmax1}) has the same value at $\qw$ and $\beta_1\qw$,
$\forall \beta_1\ne 0$, $\qw\ne 0$.
Thus, the problem of (\ref{indNotDiagSNRmax1}) is equivalent to
\al{
&\max_{\qw}\ \frac{P_s}{\sigma^2}\frac{\qw^\dagger\qR\qw}{\max_{k\in I}
\qw^\dagger\qA_k\qw}\label{indNotDiagSNRmax2}\\
&\mathrm{s.t.} \quad \qw \ne 0.\nnb
}
Similarly, the problem of (\ref{indNotDiagSNRmax2}) is equivalent to
\al{
&\max_{\qw}\ \frac{P_s}{\sigma^2}\qw^\dagger\qR\qw\label{indNotDiagSNRmax3}\\
&\mathrm{s.t.} \quad \max_{k\in I} \qw^\dagger\qA_k\qw = 1.\nnb
}
Obviously, the problem of (\ref{indNotDiagSNRmax3}) is equivalent to the problem of (\ref{indNotDiagSNRmax2f}).
This completes the proof.

\section{Calculation of $\nabla \phi_p$ and $\nabla^2 \phi_p$}\label{CalcDfD2f}

We have
\equ{
\nabla \phi_p=\frac{\partial \phi_p}{\partial \qz}=
2\sum_{k\in I} \left(\frac{\qz^T\widetilde{\qJ}_k\qz}{\phi_p(\qz)}\right)^{p-1} \widetilde{\qJ}_k\qz
}
and
\al{
\nabla^2 \phi_p&=\frac{\partial }{\partial \qz^T}\left(\frac{\partial \phi_p}{\partial \qz}\right)\nnb\\
&=2\sum_{k\in I} \left(\frac{\qz^T\widetilde{\qJ}_k\qz}{\phi_p(\qz)}\right)^{p-1} \widetilde{\qJ}_k
+\frac{1-p}{\phi_p(\qz)}
\left(\frac{\partial \phi_p}{\partial \qz}\right)\left(\frac{\partial \phi_p}{\partial \qz}\right)^T
 + \frac{4(p-1)}{\phi_p(\qz)}
\sum_{k\in I} \left(\frac{\qz^T\widetilde{\qJ}_k\qz}{\phi_p(\qz)}\right)^{p-2} \widetilde{\qJ}_k\qz\qz^T\widetilde{\qJ}_k.
}

\section{Proof of Theorem \ref{Theo:AlterMin}}\label{Proof:TheoAlterMin}

By using the Dinkelbach-type method \cite{Dinkelbach} (cf. \S\ref{Subsec:indiDiag}),
we introduce the function
\al{
F(t)=&\max_{y}\  f(t, y)\\
&\mathrm{s.t.}\ \ |y|\le \beta\nnb
}
where
\equ{
f(t, y) = a_1 |y|^2 + b_1 y + b_1^\ast y^\ast + c_1  - t(a_2 |y|^2 + b_2 y + b_2^\ast y^\ast + c_2).
}
Similarly to Property \ref{prop:F} in \S\ref{Subsec:indiDiag}, $F(t)$ is a strictly decreasing function and the equation $F(t) = 0$ has a unique root $t^\star$.
The optimal $y^\star$ associated with $F(t^\star)$ is also the solution for the problem of (\ref{indSNRmaxAlterMin}) with the optimal
objective value $t^\star$.

To obtain the expression of $F(t)$, we denote $y=|y|e^{\mathrm{i}\theta}$ and write
\al{
f(t,y) &= (a_1 - t a_2) |y|^2  + \left[(b_1 -t b_2)e^{\mathrm{i}\theta}  + (b_1 - t b_2)^\ast e^{-\mathrm{i}\theta}\right] |y| + c_1 - t c_2\nnb\\
&\le (a_1 - t a_2) |y|^2 + 2 |b_1 -t b_2||y| + c_1 - t c_2.\label{f_t_y}
}
The equality in (\ref{f_t_y}) occurs when the argument of $b_1 -t b_2$ equals $-\theta$ (if $b_1 -t b_2=0$, then $\theta$
is arbitrary).
With this, we let $r=|y|$ and write
\al{
F(t)=& c_1 - t c_2 + \max_{r}\ (a_1 - t a_2) r^2 + 2 |b_1 -t b_2|r   \\
&\quad\quad\quad\quad\ \ \mathrm{s.t.}\ \ 0\le r\le \beta\nnb.
}
Further, it is easy to get:\\
1) When $a_1 - t a_2>0$, i.e., $t < a_1/a_2$, the optimal $r$ is $\beta$ (unique), and we wirte
\equ{
F(t) = (a_1 - t a_2) \beta^2 + 2 |b_1 -t b_2|\beta  + c_1 - t c_2.\label{Ft1}
}

\noindent 2) When $a_1 - t a_2 < 0$, i.e., $t > a_1/a_2$, the optimal $r$ is given by (unique)
\equ{
r^\star = \min\left\{ \frac{|b_1-t b_2|}{t a_2 - a_1},\ \beta\right\},
}
and we write: If $t > a_1/a_2$ and $|b_1-t b_2| \ge (t a_2 - a_1)\beta$, then
\equ{
F(t) = (a_1 - t a_2) \beta^2 + 2 |b_1 -t b_2|\beta  + c_1 - t c_2;\label{Ft2a}
}
If $t > a_1/a_2$ and $|b_1-t b_2| < (t a_2 - a_1)\beta$, then
\equ{
F(t) = \frac{|b_1-t b_2|^2}{t a_2 - a_1}  + c_1 - t c_2. \label{Ft2b}
}

\noindent 3) When $a_1 - t a_2 = 0$, i.e., $t = a_1/a_2$, we know: If $b_1 -t b_2 \ne 0$ (i.e., $b_1/b_2 \ne a_1/a_2$), the optimal $r$ is $\beta$ (unique), and
\equ{
F(a_1/a_2) = 2 |b_1 - (a_1/a_2) b_2|\beta  + c_1 - (a_1/a_2) c_2; \label{Ft3a}
}
If $b_1 -t b_2 = 0$ (i.e., $b_1/b_2 = a_1/a_2$), the optimal $r$ is arbitrary in $[0, \beta]$, and
\equ{
F(a_1/a_2) = c_1 - (a_1/a_2) c_2.\label{Ft3b}
}

Recall that $F(t)$ is a strictly decreasing function and the equation $F(t) = 0$ has a unique root $t^\star$.
Thus, one and only one of the equations (\ref{Ft1}), (\ref{Ft2a}), (\ref{Ft2b}), (\ref{Ft3a}), (\ref{Ft3b})
satisfies $F(t^\star)=0$.

Based on the analysis above, it is not difficult to obtain the desired result.

\end{document}